\documentclass{aa}
\usepackage{todonotes}
\usepackage{times}
\usepackage{amssymb}
\usepackage{amsmath, bm}
\usepackage{graphicx} 
\usepackage{bmpsize}
\usepackage{epstopdf}
\usepackage{dblfloatfix} 
\graphicspath{{./image/}}
\usepackage[normalem]{ulem}
\usepackage[T1]{fontenc}
\usepackage{makecell}
\usepackage{multirow}
\usepackage{mwe}
\usepackage{mathtools}
\usepackage[colorlinks=true,allcolors=blue]{hyperref}
\usepackage{comment}

\newcommand{\HI}{\rm H{\sc i}~}

\newcommand{\TB}{\delta T_{\rm b}}
\newcommand{\MSUN}{{\rm M}_{\odot}}

\newcommand{\XHI}{x_{\rm HI}}
\newcommand{\AVXHI}{\overline{x}_{\rm HI}}
\newcommand{\AVXHISTAR}{\overline{x}_{\rm HI,\star}}
\newcommand{\AVXHIMAX}{\overline{x}_{\rm HI,min}}

\newcommand{\TS}{T_{\rm S}}

\newcommand{\TCMB}{T_{\gamma}}

\newcommand{\lya}{\rm {Ly{\alpha}}}
\newcommand{\OmegaB}{\Omega_{\rm B}}
\newcommand{\Omegam}{\Omega_{\rm m}}
\newcommand{\hmpc}{~h ~{\rm Mpc}^{-1}}
\newcommand{\DTB}{\Delta^2_{\TB}}
\newcommand{\Ddd}{\Delta^2_{\delta\delta}}
\newcommand{\Dxx}{\Delta^2_{\XHI\XHI}}
\newcommand{\Dxd}{\Delta^2_{\XHI\delta}}

\begin{document}

\title{Probing the intergalactic medium during the Epoch of Reionization using 21-cm signal power spectra}
\titlerunning{Probing the EoR IGM using 21-cm power spectra}
\author{R. Ghara\thanks{\email{ghara.raghunath@gmail.com}}\inst{1,2,3,4}, 
A. K. Shaw\thanks{\email{abinashkumarshaw@gmail.com}}\inst{2,5}, 
S. Zaroubi\inst{1,6,7},
B. Ciardi\inst{7},
G. Mellema\inst{8},
L. V. E. Koopmans\inst{1},
A. Acharya\inst{7},
M. Choudhury\inst{2,9},
S. K. Giri\inst{10},
I. T. Iliev\inst{11},
Q. Ma\inst{12}
\and
F. G. Mertens\inst{13}
}

\institute{Kapteyn Astronomical Institute, University of Groningen, PO Box 800, 9700AV Groningen, The Netherlands
\and 
ARCO (Astrophysics Research Center), Department of Natural Sciences, The Open University of Israel, 1 University Road, PO Box 808, Ra'anana 4353701, Israel
\and 
Haverford College, 370 Lancaster Ave, Haverford PA, 19041, USA
\and
Center for Particle Cosmology, Department of Physics and Astronomy, University of Pennsylvania, Philadelphia, PA 19104, USA
\and
Department of Computer Science, University of Nevada, Las Vegas, Nevada 89154, USA
\and
Department of Natural Sciences, The Open University of Israel, 1 University Road, Ra'anana 4353701, Israel
\and 
Max-Planck Institute for Astrophysics, Karl-Schwarzschild-Stra{\ss}e 1, 85748 Garching, Germany
\and
The Oskar Klein Centre, Department of Astronomy, Stockholm University, AlbaNova, SE-10691 Stockholm, Sweden
\and 
Center for Fundamental Physics of the Universe, Department of Physics, Brown University, Providence 02914, RI, USA
\and
Nordita, KTH Royal Institute of Technology and Stockholm University, Hannes Alfvéns väg 12, SE-106 91 Stockholm, Sweden
\and
Astronomy Centre, Department of Physics and Astronomy, Pevensey II Building, University of Sussex, Brighton BN1 9QH, UK
\and
School of Physics and Electronic Science, Guizhou Normal University, Guiyang 550001, PR China
\and
LERMA, Observatoire de Paris, PSL Research University, CNRS, Sorbonne Université, F-75014 Paris, France
 }

\authorrunning{Ghara et al.}

\date{Received XXX; accepted YYY}

\abstract
{The redshifted 21-cm signal from the epoch of reionization (EoR) directly probes the ionization and thermal states of the intergalactic medium during that period. In particular, the distribution of the ionized regions around the radiating sources during EoR introduces scale-dependent features in the spherically-averaged EoR 21-cm signal power spectrum. The goal is to study these scale-dependent features at different stages of reionization using numerical simulations and build a source model-independent framework to probe the properties of the intergalactic medium using EoR 21-cm signal power spectrum measurements. Under the assumption of high spin temperature, we modelled the redshift evolution of the ratio of EoR 21-cm brightness temperature power spectrum and the corresponding density power spectrum using an ansatz consisting of a set of redshift and scale-independent parameters. This set of eight parameters probes the redshift evolution of the average ionization fraction and the quantities related to the morphology of the ionized regions. We have tested this ansatz on different reionization scenarios generated using different simulation algorithms and found that it is able to recover the redshift evolution of the average neutral fraction within an absolute deviation $\lesssim0.1$. Our framework allows us to interpret 21-cm signal power spectra in terms of parameters related to the state of the IGM. This source model-independent framework is able to efficiently constrain reionization scenarios using multi-redshift power spectrum measurements with ongoing and future radio  telescopes such as LOFAR,MWA, HERA, and SKA. This will add independent information regarding the EoR IGM properties.}

\keywords{
radiative transfer - galaxies: formation - intergalactic medium - high-redshift - cosmology: theory - dark ages, reionization, first stars 
}

\maketitle

\section{INTRODUCTION}
\label{sec:intro}

The formation of the first sources of radiation at the end of the Universe's Dark Age is one of the landmark events in Cosmic history. During the first billion years, radiation from the first stars, galaxies, quasars (QSOs) and High-mass X-ray binaries (HMXBs) permanently changed the ionization and thermal state of the Universe. It is expected that radiation from early X-ray sources such as HMXBs and mini-QSOs changed the thermal state of the cold intergalactic medium (IGM) much before the IGM became highly ionized \citep[see e.g,][]{Pritchard07, Thom11, 2011MNRAS.411..955M, 2019MNRAS.487.2785I, Ross2019, 2020MNRAS.498.6083E}. The onset of the first sources that changed the IGM's thermal state is known as the `Cosmic Dawn' (CD). The subsequent period when the IGM's atomic neutral hydrogen (\HI) became ionized is known as the `Epoch of Reionization' (EoR). A few indirect probes such as the observations of Gunn-Peterson optical depth in $z\gtrsim 6$ QSO spectra and Thomson scattering optical depth of the Cosmic Microwave Background (CMB) photons provide us with useful information about the rough timing and duration of the EoR \citep[see e.g.][]{Fan06b, 2015MNRAS.447..499M, 2018Natur.553..473B,2020A&A...641A...6P, Mitra15}. However, many details about these epochs such as the exact timing, properties of the sources and their evolution, feedback mechanisms, and morphology of the ionized and heated regions are still unknown.

Observations of the redshifted 21-cm radiation produced by \HI in the IGM can provide us with information related to the timing, the morphology of the ionized and heated regions, properties of the ionizing and heating sources \citep[see e.g.][for reviews]{Pritchard12, 2013ASSL..396...45Z, Shaw_review, 2024MNRAS.530..191G}. Many of the world's large radio observation facilities have aimed for measuring the brightness temperature of this redshifted \HI 21-cm radiation (hereafter 21-cm signal) from the CD and EoR. Radio observations using single antennae such as EDGES2 \citep{EDGES2018}, SARAS2 \citep{singh2017}, REACH \citep{2022NatAs...6..984D} and LEDA \citep{price2018} aim to measure the redshift evolution of the sky-averaged 21-cm signal. However, observing the morphological distribution of the 21-cm signal in the sky is expected to tell us more about these epochs. Radio interferometers such as the Low-Frequency Array (LOFAR)\footnote{\url{http://www.lofar.org/}} \citep{vanHaarlem2013LOFAR:ARray, 2017ApJ...838...65P}, the New Extension in Nançay Upgrading LOFAR (NenuFAR)\footnote{ \url{https://nenufar.obs-nancay.fr/en/homepage-en/}}\citep{refId0}, the Amsterdam ASTRON Radio Transients Facility And Analysis Center (AARTFAAC)\citep{2022A&A...662A..97G},  the Precision Array for Probing the Epoch of Reionization (PAPER)\footnote{\url{http://eor.berkeley.edu/}} \citep{parsons13, 2019ApJ...883..133K}, the Murchison Widefield Array (MWA)\footnote{\url{http://www.mwatelescope.org/}} \citep[e.g.][]{tingay13, Wayth2018mwa} and the Hydrogen Epoch of Reionization Array (HERA)\footnote{\url{https://reionization.org/}} \citep{2017PASP..129d5001D} have been commissioned to measure the spatial fluctuations in the \HI 21-cm signal at different stages of the CD and EoR. 

Due to limited sensitivity, the radio interferometer-based observations aim to detect this signal in terms of the statistical quantities such as the spherically-averaged power spectrum ($\DTB(k,z)$) of the differential brightness temperature ($\TB$) of the \HI signal at different redshifts ($z$) and scales/wave-numbers ($k$). The upcoming Square Kilometre Array (SKA)\footnote{\url{http://www.skatelescope.org/}} will be more sensitive and will also produce tomographic images of the CD and EoR 21-cm signal \citep{2015aska.confE..10M, ghara16}.

Observing the 21-cm signal from CD and EoR is very challenging and it has remained undetected by the radio observations to date. The measured \HI signal is severely contaminated by the galactic and extra-galactic foregrounds. While the foregrounds are more substantial than the expected CD and EoR \HI signal by several orders of magnitude \citep[see e.g.,][]{ghosh12}, their smooth frequency dependence allows them to be either subtracted \citep{2009MNRAS.397.1138H, 2015MNRAS.447.1973B, 2016MNRAS.458.2928C, 2018MNRAS.478.3640M, 2021MNRAS.500.2264H}, avoided \citep{2010ApJ...724..526D, 2014PhRvD..90b3019L} or suppressed \citep{kanan2007MNRAS.382..809D, ghara15c}. These observations also face severe challenges at the calibration step of the data analysis process. Nevertheless, recent improvements in the calibration methods \citep[see e.g.,][]{2019ApJ...884..105K, 2020ApJ...888...70K, 2020Mevius, Gan2022, 2023A&A...669A..20G}, and the mitigation of the foregrounds \citep[e.g.,][]{2018MNRAS.478.3640M, 2014PhRvD..90b3019L}  made it possible to obtain noise dominated upper limits of $\DTB(k,z)$. For example,  $\DTB(k=0.14 \hmpc,~z=6.5)\approx (43)^2 ~{\rm mK}^2$, $\DTB(k = 0.075 \hmpc, ~z=9.1) \approx (73)^2 ~{\rm mK}^2$ and $\DTB(k = 0.34 \hmpc, ~z=7.9) \approx {(21.4)}^{2} ~{\rm mK}^2$ are the best upper limits obtained from MWA \citep{2020MNRAS.493.4711T}, LOFAR \citep{2020MNRAS.493.1662M} and HERA \citep{Abdurashidova_2023} EoR observations, respectively. 
 
 These recent upper limits have started to rule out CD and EoR scenarios including those which do not require either an unconventional cooling mechanism or the presence of a strong radio background in addition to the CMB \citep[e.g.,][]{2020MNRAS.493.4728G, 2020arXiv200603203G, 2020MNRAS.498.4178M, 2022ApJ...924...51A}. For example, the recent HERA EoR observation results as reported in \citet{2022ApJ...925..221A}, \citet{2022ApJ...924...51A} show that the IGM temperature must be larger than the adiabatic cooling threshold by redshift 8 while the soft band X-ray luminosity per star formation rate of the first galaxies are constrained ($1\sigma$ level) to [$10^{40.2}-10^{41.9}$] erg/s/($\MSUN$/yr). In addition, the recent results from the global \HI 21-cm signal observations such as SARAS and EDGES have also started ruling out EoR and CD scenarios and putting constraints on the properties of the early sources, models of dark matter and level of radio backgrounds \citep[e.g.,][]{2018Natur.555...71B, 2018PhRvL.121a1101F, 2018Natur.557..684M, 2019JCAP...04..051N, 2019MNRAS.487.3560C, 2020MNRAS.492..634G, 2020MNRAS.496.1445C, 2022JCAP...03..055G, 2023JApA...44...10B}.

\begin{figure*}
\begin{center}
\includegraphics[scale=0.75]{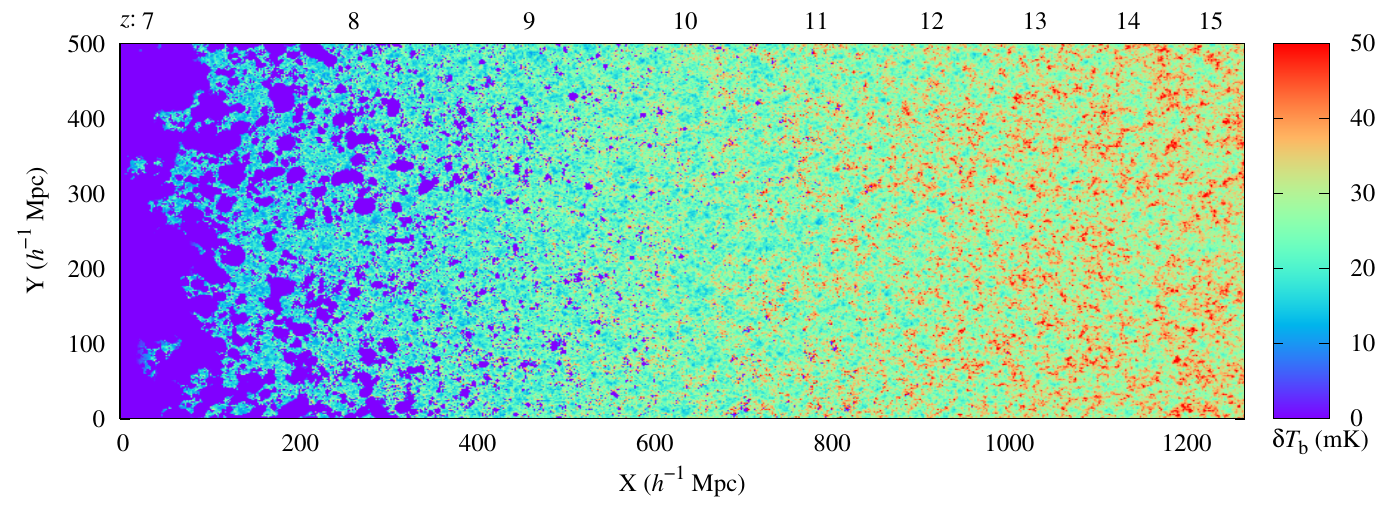}
    \caption{Light-cone of EoR 21-cm signal. This shows the redshift ($z$) evolution of the corresponding differential brightness temperature ($\TB$) from $z \approx 15.5$ to $6.9$. We assume a high spin temperature limit, i.e., $\TS \gg \TCMB$. This light cone is generated using the {\sc grizzly} code and represents our fiducial EoR scenario.}
   \label{image_tbslice}
\end{center}
\end{figure*}

These previous studies have put constraints mainly on the astrophysical source properties using either Bayesian inference techniques \citep[e.g.,][]{Park2019InferringSignal, 2020MNRAS.495.4845C} or Fisher matrices \citep[e.g.,][]{2016MNRAS.458.2710E, 2020MNRAS.498.1480S}. The main reason behind this is the fact that 21-cm signal simulation codes take the source parameters as input. However, it should be realized that the 21-cm signal measurements do not probe the astrophysical sources directly. In addition, the inference on the properties of the astrophysical sources is limited by the ambiguity of the source model used in the inference framework. The observed 21-cm signal, on the other hand, directly probes the ionization and the thermal states of the IGM. Therefore, we emphatically aim to constrain the IGM properties rather than the astrophysical source parameters.

Previously, \citet{2013ApJ...777..118M} considered the features of the redshift evolution of the sky-averaged brightness temperature curves within a simplified global \HI signal framework which does not invoke any astrophysical sources and attempted to constrain physical properties of the IGM in terms of $\lya$ background, overall heat deposition, mean ionization fraction, and their time derivatives. In the context of 21-cm signal power spectrum, studies such as \citet[][]{2020MNRAS.493.4728G, 2021MNRAS.503.4551G} used the recently obtained upper limits from LOFAR \citep{2020MNRAS.493.1662M} and MWA \citep{2020MNRAS.493.4711T} to constrain the properties of the IGM at different stages of the EoR. These studies use the outputs from {\sc grizzly}  \citep{ghara15a} simulations and characterise the IGM in terms of quantities such as the sky-averaged ionization fraction, average gas temperature, sky-averaged brightness temperature, the volume fraction of the `heated regions’ in the IGM with its brightness temperature $T_{\rm b}$ larger than the background CMB temperature $\TCMB$, the characteristic size of these heated regions. For example, using the recent upper limits from LOFAR \citep{2020MNRAS.493.1662M}, \citet{2020MNRAS.493.4728G} ruled out reionization scenarios at redshift 9.1 where heating of the gas is negligible and the IGM is characterised by ionized fraction $\gtrsim 0.13$, a distribution of the ionized regions with a characteristic size $\gtrsim 8 ~h^{-1}~ \rm Mpc$, and a full width at half-maximum $\gtrsim 16 ~h^{-1}~\rm Mpc$. In an alternative approach, \citet{2022RAA....22c5027S} used Artificial Neural Networks to build a framework that estimates the size distribution of the ionized regions using the EoR 21-cm power spectrum.

Our previous studies such as \citet[][]{2020MNRAS.493.4728G} and \citet[][]{2021MNRAS.503.4551G}, which aim at constraining the properties of the CD and EoR IGM parameters, use a source-parameter dependent {\sc grizzly} simulations. The inputs of their framework are a set of source parameters such as the ionization efficiency, the minimum mass of dark matter halos that host UV emitting sources, the X-ray emission efficiency, the minimum mass of dark matter halos that host X-ray emitting sources. The framework provides a set of derived IGM parameters in addition to the 21-cm signal observable. It is not straightforward to build a mathematical framework that directly connects the complex morphology of the IGM to the 21-cm signal observable by skipping the source-parameter dependence.  Recently, \citet{2022MNRAS.514.2010M} have attempted to build such a galaxy-free phenomenological model for the EoR 21-cm signal power spectra. The model assumes uniform $\TS$, spherical ionized bubbles and binary ionization field. While the model efficiently predicts the 21-cm signal power spectrum for volume average neutral fraction $\AVXHI$ $\gtrsim 0.8$, the prediction accuracy rapidly drops for reionization stages with  $\AVXHI$ $\lesssim 0.8$ which shows the complexity level of the problem.

Unlike our aforementioned IGM inference framework, the main goal of this work is to develop a source parameter-free phenomenological model of EoR 21-cm signal power spectra in terms of quantities related to the IGM. We keep our model simple by ignoring the effect of spin-temperature fluctuations and targeting the IGM only during the EoR. The amplitude and the shape of $\DTB(k,z)$ as a function of $k$ during different stages of the EoR depend on the ionization fraction and the complex morphology of the ionized regions at that period. The aim here is to use the multi-redshift measurements of the EoR 21-cm signal power spectra to constrain the IGM properties during the EoR.

This paper is structured as follows. In Section \ref{sec:method}, we describe the basic methodology of our framework.  We present our results in Section \ref{sec:results}, before concluding in Section \ref{sec:con}. The cosmological parameters used throughout this study are the same as the $N$-body simulations employed here, i.e.  $\Omegam=0.27$, $\Omega_\Lambda=0.73$, $\OmegaB=0.044$, $h=0.7$  \citep[Wilkinson ~Microwave ~Anisotropy ~Probe (WMAP);][]{2013ApJS..208...19H}.

\section{Framework}
\label{sec:method}

\subsection{The EoR 21-cm signal}
\label{sec:psmodelXHI}

 The differential brightness temperature ($\TB$) of the 21-cm signal from a region at angular position $\vec{x}$ and redshift $z$  can be expressed as  \citep[see e.g.,][]{madau1997, Furlanetto2006},
\begin{align}
 \TB (\vec{x}, z)  & = 27 ~ \XHI (\vec{x}, z) [1+\delta_{\rm B}(\vec{x}, z)] \left(\frac{\OmegaB h^2}{0.023}\right) \nonumber\\
&\times \left(\frac{0.15}{\Omegam h^2}\frac{1+z}{10}\right)^{1/2}\left[1-\frac{\TCMB(z)}{\TS(\vec{x}, z)}\right]\,\rm{mK}.
\nonumber \\
\label{eq:brightnessT}
\end{align}
Here, $\TS$, $x_{\rm HI}$ and $\delta_{\rm B}$ are respectively the spin temperature of \HI, the neutral hydrogen fraction and the baryonic density contrast of the region located at $(\vec{x}, z)$. The quantity $\TCMB$ is the radio background temperature at 21-cm wavelength for redshift $z$. In this study, we assume a high spin temperature limit, i.e., $\TS \gg \TCMB$. This is expected to be the case in the presence of efficient X-ray heating.

\begin{figure}
\begin{center}
\includegraphics[scale=0.54]{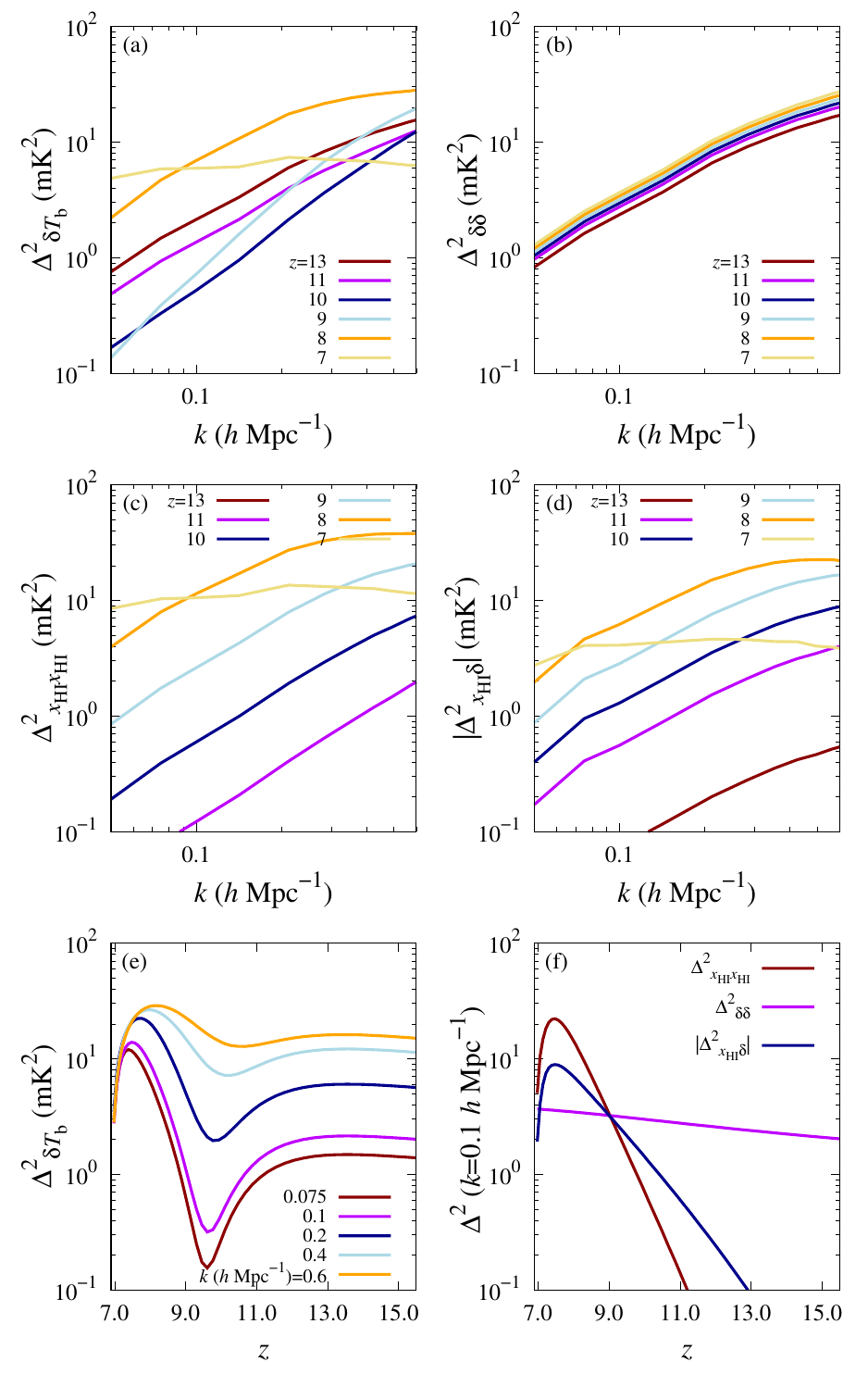}
    \caption{Power spectra of EoR 21-cm signal brightness temperature and their different components. The (a), (b), (c) and (d) panels show $\DTB, ~\Ddd$, $\Dxx$ and $|\Dxd|$ respectively as a function of $k$ at different stages of reionization. The (e) panel shows the redshift evolution of the 21-cm power spectrum at different scales. The (f) panel compares the redshift evolution of $\Dxx, ~\Ddd$ and $|\Dxd|$ at $k=0.1 \hmpc$.   The power spectra correspond to our fiducial {\sc grizzly} EoR scenario as shown in Figure \ref{image_tbslice}.}  
   \label{image_psfid}
\end{center}
\end{figure}

Here, we use the {\sc grizzly} code \citep{ghara15a, ghara15b} to generate brightness temperature maps during the EoR. The inputs for this code are the uniformly gridded dark-matter density and velocity field cubes and the corresponding dark-matter halo list. The {\sc grizzly} simulations considered in this study use the dark-matter fields and the corresponding halo lists within comoving cubes of side $500 ~h^{-1}$ Mpc, produced from the PRACE\footnote{Partnership for Advanced Computing in Europe: \url{http://www.prace-ri.eu/}} project PRACE4LOFAR \textit{N}-body simulations \citep[see e.g,][for the details of the simulation]{2019JCAP...02..058G, 2021MNRAS.502.3800K}. The simulation assumes that all dark-matter halos with masses larger than $M_{\rm min}$ contribute to reionization. The stellar mass $M_\star$ inside a halo of mass $M_{\rm halo}$ is assumed to be $M_\star \propto M^{\alpha_s}_{\rm halo}$.  We choose the ionization efficiency ($\zeta$) so that the reionization process ends roughly at $z\sim 6.5$\footnote{Some probes such as the $\lya$ forest observations at  $z\approx5.5$ suggest a late reionization compared to our fiducial reionization model \citep[see e.g.,][]{2018ApJ...863...92B, 2018ApJ...864...53E}. However, the exact end of the EoR is still debated and hence, we choose our fiducial simulation from our earlier works \citep[e.g.,][]{2023MNRAS.522.2188S, 2024MNRAS.530..191G}.}. The reionization models considered in this study are inside-out in nature where the very dense regions around the sources get ionized first. We refer the reader to \citet{ghara15a, 2020MNRAS.493.4728G} for the details of the method and the source parameters. Our fiducial {\sc grizzly} model, as shown in Figure \ref{image_tbslice}, corresponds to a choice of $M_{\rm min}=10^9 ~\MSUN$ and $\alpha_s=1$ and spans from redshift $6.9$ to $15.5$. We also produce 23 more reionization scenarios by choosing different combinations of [$M_{\rm min}, \alpha_s$] where we vary $M_{\rm min}$ between $10^9-10^{11} ~\MSUN$ and $\alpha_s$ between $0.3-2$. Smaller values of $M_{\rm min}$ and larger values of $\alpha_s$ will create a more patchy reionization scenario. We use all these reionization scenarios for building and testing our model of the $\TB$ power spectrum. Note that all our simulations include the redshift-space distortion effects based on the cell moving method \citep{ghara15a,2021MNRAS.506.3717R}. 

Figure \ref{image_tbslice} shows a slice through a simulated light-cone of the EoR 21-cm signal $\TB$ (Equation \ref{eq:brightnessT}). The figure represents how the fluctuations $\TB$ in the sky (shown by the vertical axis) evolve with redshift/distance from the observer (represented by the horizontal axis). Our assumption of $\TS \gg \TCMB$ makes the \HI 21-cm signal $\TB$ positive in the neutral regions while the ionized regions are represented by $\TB=0$. Note that ionized regions in the IGM are absent around $z=15$ where the fluctuations in $\TB$ are governed by the density fluctuations only (Equation \ref{eq:brightnessT}). Small isolated ionized regions gradually appear around the high-density peaks in $\delta_{\rm B}$. Over time, the isolated ionized regions grow in size and overlap with each other. This overlap can occur as early as when the IGM volume is ionized by a few tens of percent depending upon the reionization history. These overlaps eventually create complex percolated structures of the ionized regions which grow in volume over time as the reionization progresses. For $\AVXHI\gtrsim 0.5$ (i.e. $z \gtrsim 8$ in Figure \ref{image_tbslice}), the sizes of the ionized regions are smaller than the neutral regions. Visually, the ionized regions are embedded into the neutral regions. It becomes the opposite at reionization stages with $\AVXHI\lesssim 0.5$. At these stages, the distribution of the neutral regions is more meaningful compared to the distribution of the ionized regions.

\subsection{The EoR 21-cm signal power spectrum}
\label{sec:eorps}

This study is based on the $k$-dependent features of the dimensionless power spectrum of coeval $\TB$ cube at redshift $z$, i.e., $\DTB(z, k) = k^3 P_{\TB}(z, k)/2\pi^2$ during different stages of the EoR. Assuming statistical homogeneity of the signal, one can define the 3D power spectrum for a coeval signal volume $V$ as $ \delta_{\rm K}(\vec{k} - \vec{k}^\prime) P_{\TB}(z, \vec{k}) = V^{-1} \langle \hat{\TB}(z, \vec{k}) ~{\hat{\TB}}^{*}(z, \vec{k}^\prime)\rangle$, where $\delta_{\rm K}(\vec{k} - \vec{k}^\prime)$ denotes the 3D Kroneker's delta function and $\hat{\TB}(z, \vec{k})$ is the Fourier transform of the EoR 21-cm signal $\TB(\vec{x}, z)$. Here we use the spherically-averaged power spectrum $P_{\TB}(z, k)$ which is computed by averaging $P_{\TB}(z, \vec{k})$ within spherical shells of certain widths in 3D Fourier space. According to Equation (\ref{eq:brightnessT}), the EoR 21-cm power spectrum $\DTB(z, k)$ depends on the power spectra of the density and neutral fraction fields ($\Ddd$ and $\Dxx$ respectively), and their cross power spectrum $\Dxd$. Here, $\Ddd$ is associated with a field given by Equation (\ref{eq:brightnessT}) for $x_{\rm HI} (\vec{x}, z)=1$ , therefore powered by the density fluctuations only. On the other hand, the field associated with $\Dxx$ assumes $\delta_{\rm B}(\vec{x}, z)=0$ in Equation (\ref{eq:brightnessT}), and thus is independent of the density fluctuations and only depends on the neutral fraction fluctuations  \citep[][]{2007ApJ...659..865L, 2022MNRAS.513.5109G}\footnote{Note that \citet[][]{2007ApJ...659..865L, 2022MNRAS.513.5109G} use the field of fluctuations in $\XHI$, $\delta \XHI$. This leads to additional, higher-order, cross terms when composing the 21-cm power spectra in terms of the constituent fields $\delta$ and $\delta \XHI$.}. $\Dxd$ is the cross-power spectrum of the fields associated with $\Ddd$ and $\Dxx$.

 Figure \ref{image_psfid} shows the evolution of the EoR 21-cm signal power spectrum $\DTB$ as well as the power spectra of the density field $\Ddd$, neutral fraction field $\Dxx$, and their cross-power spectrum $\Dxd$. The power spectra correspond to our fiducial {\sc grizzly} EoR scenario as presented in Figure \ref{image_tbslice}. The $a$, $b$, $c$ and $d$ panels of Figure \ref{image_psfid} show $\DTB$, $\Ddd$, $\Dxx$ and $|\Dxd|$ respectively as a function of $k$ at different redshifts. The panel $e$ shows the redshift evolution of $\DTB$ for different scales while panel $f$ compares the redshift evolution of the $\Dxx, ~\Ddd$ and $|\Dxd|$ for $k =0.1 \hmpc$. The high-density regions get ionized first in this inside-out reionization model. This causes anti-correlation between $\XHI$ and $\delta$ and thus, negative values for the cross-power spectrum $\Dxd$ which suppresses the large-scale $\TB$ power spectrum at the initial stage of the EoR \footnote{This phase of strong suppression of the large scale 21-cm power spectrum was first pointed out by \citet[][]{2007ApJ...659..865L} who called it ``equilibration''. \citet[][]{2022MNRAS.513.5109G} studied it in some more detail and showed that it is caused by the near cancellation of the positive and negative terms in the decomposition of the 21-cm power spectra.}. However, the suppression is less significant at the small-scales, causing a tilt in the $\DTB$ compared to the $\Ddd$ (see panels $a$ and $b$). For $z\gtrsim 9$, $|\Dxd|$ remains larger than $\Dxx$ (see panels $d$ and $f$). For $z\lesssim 9$, $\Dxx$ becomes the dominant term. The interplay between the $\Dxx$ and $\Dxd$ contributions causes a minimum in the $\DTB$ vs $z$ curves around $z\sim 10$ (see bottom panels). The large-scale $\TB$ power spectrum increases as reionization progresses. For example, $\DTB(k=0.1 \hmpc)$ increases from $z=9$ to $z=7.3$ as $\Dxx$ becomes dominant compared to $\Dxd$. For $z\lesssim 7.3$, $\DTB(k=0.1 \hmpc)$ quickly drops as the majority of the IGM gets ionized. This causes $\DTB(k=0.1 \hmpc)$ to peak at redshift 7.3 with amplitude of $\approx 10 ~\rm mK^2$. The peak amplitudes and the associated redshifts change with $k$ (see panel $e$ of Figure \ref{image_psfid}).

\begin{figure}
\begin{center}
\includegraphics[scale=1.]{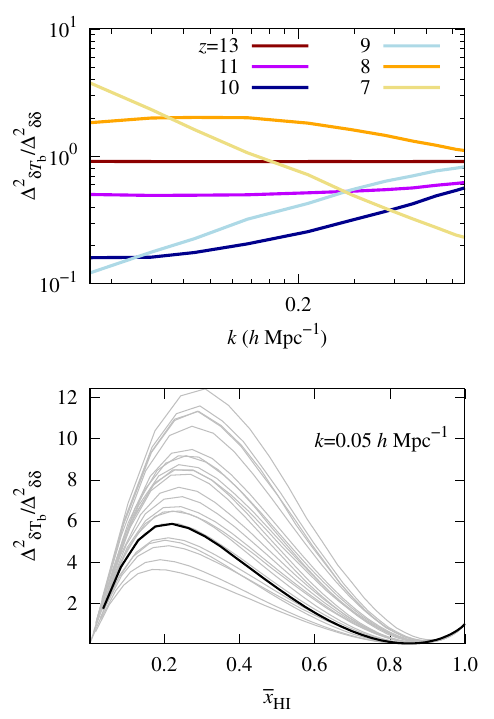}
    \caption{Ratio of $\DTB$ and $\Ddd$ obtained from simulations during EoR. The top panel shows  $\DTB/\Ddd$ as a function of $k$ at different redshifts. These correspond to our fiducial {\sc grizzly} model which is also shown in Figure \ref{image_psfid}. The bottom panel shows evolution of $\DTB/\Ddd$ at $k=0.05 \hmpc$ as a function of neutral fraction. Different curves stand for different reionization scenarios generated using {\sc grizzly} by varying $\alpha_S$ and $M_{\rm min}$. The black curve represents our fiducial {\sc grizzly} simulation which corresponds to $\alpha_S=1$ and $M_{\rm min}=10^9 ~\MSUN$.}  
   \label{image_psevo}
\end{center}
\end{figure}

The top panel of Figure \ref{image_psevo} shows the ratio of $\DTB$ and $\Ddd$ (also known as the 21-cm signal bias) as a function of $k$ at different redshifts for the fiducial {\sc grizzly} model.  The bottom panel of Figure \ref{image_psevo} shows the evolution of $\DTB/\Ddd$ for $k=0.05 \hmpc$ as a function of $\AVXHI$. The different curves in the bottom panels correspond to different {\sc grizzly} reionization models, with the thick black curve representing the fiducial one.  This ratio is expected to be $1$ for $\AVXHI=1$. The curves show that the ratio first decreases from $1$ to a minimum at an early stage of reionization. We denote $\AVXHI$ at this stage as $\AVXHIMAX$. This minimum corresponds to the equilibration phase caused by the anti-correlation between the density and neutral fraction fields in our inside-out reionization model where $\Dxd<0$. The ratio then increases with the increase of the size of the ionized regions and reaches a maximum and further decreases to zero as $\AVXHI$ approaches zero towards the end of the EoR. We denote the ionization fraction at the stage when the maximum occurs as $\AVXHISTAR$. The qualitative features of the different curves in the bottom panel of Figure \ref{image_psevo} are similar, although the stages when the minimum and maximum occur (i.e., the values of $\AVXHIMAX$ and $\AVXHISTAR$) and the peak amplitude of the ratio $\DTB/ \Ddd$ changes with the patchiness of the reionization scenarios. The curve which reaches the largest bias or ratio value (approximately 14) corresponds to the model which uses the largest values for both $M_{\rm min}$ ($10^{11} \MSUN$) and $\alpha_S$ (2), and thus represents the most patchy reionization scenario among the considered models. Nevertheless, the evolutionary features of this ratio as a function of $\AVXHI$ remain the same despite being an early/late or fast/slow reionization.

\begin{figure*}
\begin{center}
\includegraphics[scale=0.4]{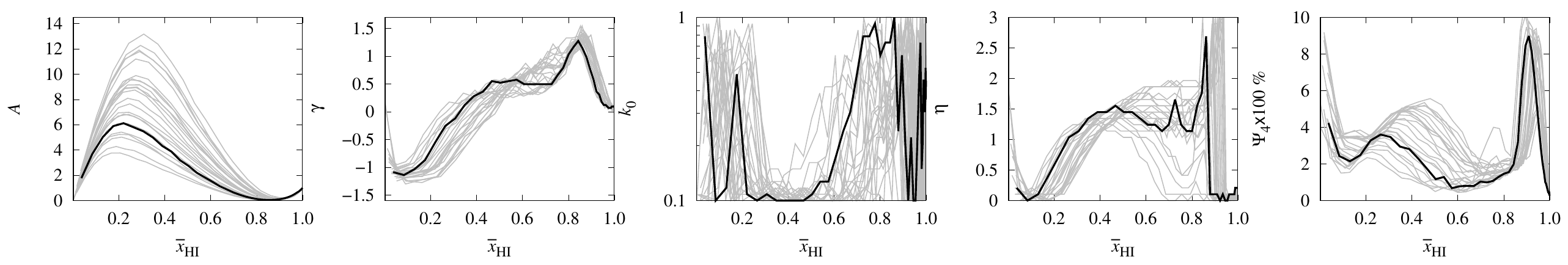}
    \caption{Outcome of fitting EoR power spectra using Equation (\ref{eq.deltb1}). Left to right panels show the evolution of the best-fit values of the parameters $A$, $\gamma$, $k_0$ and $\eta$ and fitting error as a function of $\AVXHI$ for different reionization scenarios generated using {\sc grizzly}. Bold lines correspond to the fiducial {\sc grizzly} reionization scenario.}  
   \label{image_psfit}
\end{center}
\end{figure*}

\begin{figure*}
\begin{center}
\includegraphics[scale=0.67]{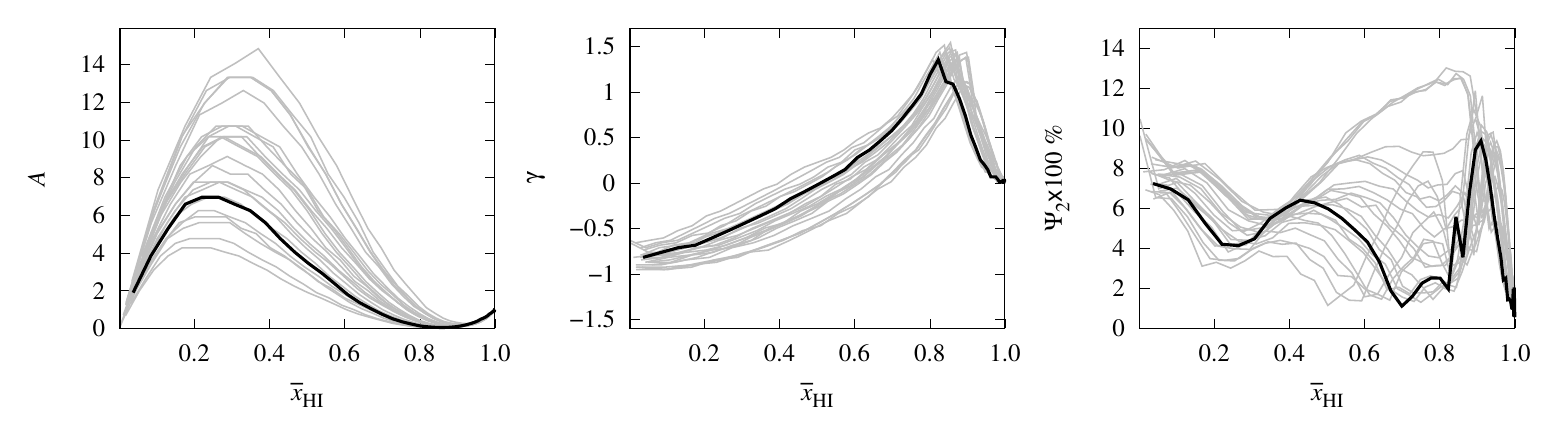}
    \caption{Outcome of fitting EoR power spectra using Equation (\ref{Equ.par1}). Left to right panels show the evolution of the best-fit values of the parameters $A$ and $\gamma$ and fitting error as a function of $\AVXHI$ for different reionization scenarios. Note that here we have used Equation (\ref{Equ.par1}) while we used Equation (\ref{eq.deltb1}) for Figure \ref{image_psfit}.   The reionization scenarios are the same as in Figure \ref{image_psfit}. Bold lines correspond to the fiducial {\sc grizzly} reionization scenario.}
  \label{image_psfit2}
\end{center}
\end{figure*}

\subsection{Modelling scale dependence of the EoR 21-cm signal power spectrum at a given redshift}
\label{sec:psmodel}
In this section, we aim to model the complex scale dependence of the $\TB$ power spectrum (e.g., see Figure \ref{image_psfid} and \ref{image_psevo}) as we described in the previous section. The $k$-dependence of $\DTB$ evolves with time/redshift. Note that this study considers features of the power spectrum for the range $0.05 \hmpc \lesssim k \lesssim 0.6 \hmpc$, which covers the scales probed by EoR observations such as LOFAR, MWA.  The overall feature of the power spectrum, as we have seen in the top panel of Figure \ref{image_psevo} \citep[see also,][]{10.1093/mnras/stz2926, 2022MNRAS.513.5109G}, suggests that one possible ansatz to represent the $k$-dependence of the coeval EoR 21-cm signal power spectrum can be
\begin{equation}
\DTB(k,z)=A~\frac{\left(\frac{k}{k_c}\right)^\gamma}{1+\left(\frac{k}{k_0}\right)^\eta}~\Ddd(k,z).
\label{eq.deltb1}
\end{equation}
Here $A, k_c, k_0, \gamma, \eta$ are the parameters to fit $\DTB$ at a particular redshift, considering $\Ddd$ to be known for the background cosmology. The form of the numerator in Equation (\ref{eq.deltb1}) is chosen to compensate for the difference in slope between $\DTB$ and $\Ddd$ during the early stages of the EoR (see Figure \ref{image_psfid}). On the other hand, the denominator accounts for the fall of $\DTB$ relative to the corresponding $\Ddd$ at the small scales (e.g., for $k\gtrsim 0.1 \hmpc$) during the advanced stages of the EoR (e.g., for $\AVXHI\lesssim 0.7$). For $\AVXHI=1$, one expects $A=1, \gamma=0$, $k_0\rightarrow \infty$ and $\eta$ to be positive. It is expected that the $k_0$ values are much larger compared to the smallest $k$ value achieved by the EoR \HI observations. For example, LOFAR reaches $k\sim 0.05 \hmpc$ as reported in papers such as \citet{2017ApJ...838...65P}. We set $k_c=0.05 \hmpc$ which makes the $A$ parameter equal to the ratio of $\DTB$ and $\Ddd$ at $k= 0.05 \hmpc$ if $k_0$ is quite large compared to $0.05 \hmpc$.

We first attempt to fit the power spectra at different redshifts using Equation \ref{eq.deltb1}, separately, by varying $A, ~k_0,~ \gamma,~{\rm and}~ \eta$ on a grid for our fiducial {\sc grizzly} scenario. We explore the $\log(A)$, $\gamma$, $\log(k_0)$ and $\eta$ parameter space on equal-spaced grids. The chosen parameter ranges are $[-4,2]$, $[-2,2]$, $[-1,0]$ and $[0, 3]$, respectively. We estimate the best-fit values of these four parameters at each redshift by minimising the error 
\begin{equation}
\Psi^2_4(\mathcal{S}:[\small{z, A, \gamma, k_0, \eta}]) = \frac{1}{n}\sum_{i=1,n} \left({\frac{\Delta^2_{\TB,m}(\mathcal{S})}{\Delta^2_{\TB,o}(z, k_i)}-1}\right)^2. \nonumber
\end{equation}
We consider a $k$ range in between 0.05 and  $ 0.6 ~h ~\rm Mpc^{-1}$ and divide it into $n=9$ log spaced bins. Here, $\Delta^2_{\TB,m}$ represents the $modelled$ power spectrum using Equation (\ref{eq.deltb1})  for a set of input parameters.  $\Delta^2_{\TB,o}$ is the $observed/input$ power spectrum which we consider for fitting. The evolution of the best-fit parameter values as a function of $\AVXHI$ is shown in the different panels of Figure \ref{image_psfit}. The rightmost panel, which shows the goodness of the fit, indicates that the maximum error in our fitting is well within $10\%$.

The left panel of Figure \ref{image_psfit} presents the best-fit values of $A$ as a function of $\AVXHI$. These roughly agree with the values presented in Figure \ref{image_psevo}. We find that $k_0$ reaches the maximum value of the range or becomes unconstrained (simultaneously $\eta$ becomes ill-defined) during the early stages of reionization ($\AVXHI \geq \AVXHIMAX$) before significant overlap between the isolated ionized regions occurs. Eventually, the formation of large overlapped ionized regions changes the $k$ dependencies on the small-scale power spectra. With the growth of overlapping ionized regions, the characteristic bubble size increases, which implies a decrease in $k_0$ values while remaining much larger than $ 0.05 \hmpc$. We repeated the fitting of Equation (\ref{eq.deltb1}) for our different {\sc grizzly} scenarios and found a qualitatively similar dependence of $k_0$ and $\eta$ on reionization history (see grey lines in Figure \ref{image_psfit}).

It is expected from Equation \ref{eq.deltb1} that the parameters $\gamma, k_0$ and $\eta$ might be degenerate. To reduce degenerate parameters and to minimise the number of parameters in our ansatz,  we fix $k_0$ and $\eta$ at typical values of the best-fit parameters throughout the reionization history. We fixed $\eta=1.5$,  while we fixed $k_0=0.3 \hmpc$ for $\AVXHI\lesssim \AVXHIMAX$ and infinity otherwise. This reduces the number of parameters and modifies Equation (\ref{eq.deltb1}) to
\begin{equation}
\DTB = \begin{dcases*} \Ddd A \frac{\left(\frac{k}{0.05}\right)^\gamma}{1+\left(\frac{k}{0.3}\right)^{1.5}},~ & $\rm if ~ \AVXHI\lesssim \AVXHIMAX$.\\
  \Ddd A \left(\frac{k}{0.05}\right)^\gamma, & otherwise. \\
          \end{dcases*}
\label{Equ.par1}
\end{equation}
As before, we fixed $k_c$ to 0.05, which ensures that the parameter $A$ represents the ratio $\DTB/\Ddd$ at $k= 0.05 \hmpc$.

To check the performance of the form in Equation (\ref{Equ.par1}), we vary the parameters $A$ and $\gamma$ and compare the fitted power spectrum $\Delta^2_{\TB,m}(z, k, A, \gamma)$ with the simulated input power spectrum $\Delta^2_{\TB,o}(z, k)$ at each redshift independently. We explore the $\log(A)$ and $\gamma$ parameter space on grids where we have chosen the same parameter ranges as above, i.e. $[-4,2]$ and $[-2,2]$, respectively. We estimate the best-fit values of $A$ and $\gamma$ at each redshifts by minimising the error 
\begin{equation}
\Psi^2_2(z, A, \gamma) = \frac{1}{n}\sum_{i=1,n} \left({\frac{\Delta^2_{\TB,m}(z, k_i, A, \gamma)}{\Delta^2_{\TB,o}(z, k_i)}-1}\right)^2. \nonumber
\end{equation}
The left two panels of Figure \ref{image_psfit2} present the evolution of the best-fit values of $A$ and $\gamma$ as a function of $\AVXHI$ for different {\sc grizzly} reionization scenarios. The right panel of the figure presents $\Psi_2\times 100$ as a function of $\AVXHI$. The figure shows that, even with the simplified form of $\DTB$ as used in Equation (\ref{Equ.par1}), the fitting error $\Psi_2$ is  $\lesssim 10$ percent and not drastically different compared to $\Psi_4$. On the other hand, the evolution of the $\gamma$ parameter is now smoother compared to the four-parameter case.

\begin{figure}
\begin{center}
\includegraphics[scale=0.7]{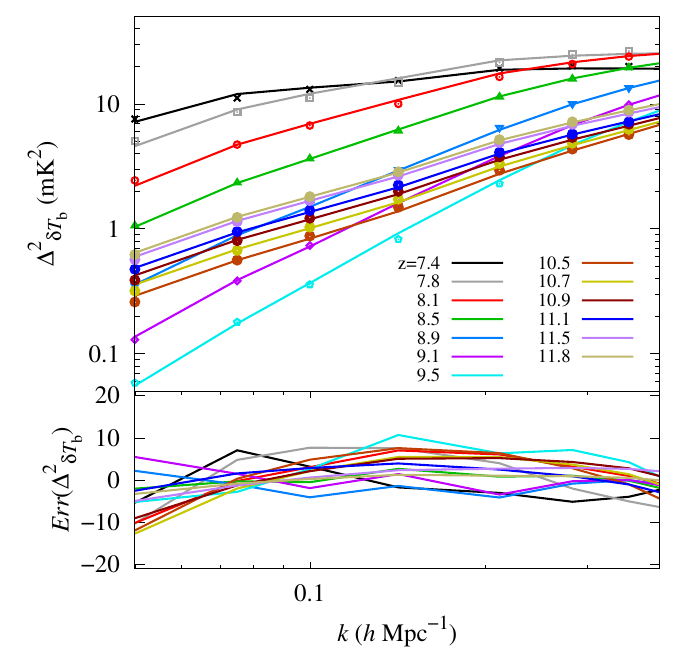}
    \caption{Comparing simulated and fitted EoR power spectra using Equation (\ref{Equ.par1}). The top panel shows a comparison between the 21-cm signal power spectrum of our fiducial reionization scenario and its best-fit power spectrum obtained using Equation (\ref{Equ.par1}). Different colors represent different redshifts. The lines correspond to the power spectrum $\DTB$ from {\sc grizzly} simulations while the points stand for the fitted power spectrum $\Delta^2_{\TB, \rm f}$. The bottom panel shows the percentage fitting error $Err(\DTB)=(1-\frac{\Delta^2_{\TB, \rm f}}{\DTB})\times 100\%$.}
   \label{image_perrorsfit}
\end{center}
\end{figure}

The top panel of  Figure \ref{image_perrorsfit} shows a comparison between the power spectrum of our fiducial reionization scenario and its best-fit power spectrum $\Delta^2_{\TB,\rm f}(z, k)$ obtained using Equation (\ref{Equ.par1}). The bottom panel of Figure \ref{image_perrorsfit} shows the percentage fitting error $Err(\DTB) = (1-\frac{\Delta^2_{\TB, \rm f}}{\DTB})\times 100 \%$ as a function of $k$ modes for the different redshifts. The plot shows that Equation (\ref{Equ.par1}) can predict the power spectrum of the fiducial EoR scenario with error $\lesssim 10\%$.  We roughly find similar results for other {\sc grizzly} reionization scenarios.

\subsection{Modelling the redshift evolution of the EoR 21-cm signal power spectrum}
\label{sec:psmodelz}
We already find that the form of the 21-cm power spectrum as used in Equation (\ref{Equ.par1}) works well for different stages of reionization. However, for any given reionization scenario, the best-fit values of the $A$ and $\gamma$ evolve with redshift or alternatively $\AVXHI$. Thus, it is important to understand the behaviour of $A$ and $\gamma$ parameters as a function of $\AVXHI$ to come up with a final set of redshift-independent parameters which can be constrained using multi-redshift EoR 21-cm signal power spectra measurements.

The left panel of Figure \ref{image_psfit2} shows the dependencies of the best-fit values of $A$ as a function of $\AVXHI$. Note that both the bottom panel of Figure \ref{image_psevo} and the left panel of Figure \ref{image_psfit2} represent the evolution of $\DTB/\Ddd$ at $k=0.05 \hmpc$ as a function of $\AVXHI$. While the curves in the bottom panel of Figure \ref{image_psevo} show the estimates directly from the {\sc grizzly} simulation, the curves in the left panel of Figure \ref{image_psfit2} represent the best-fit value of parameter $A$ when using Equation (\ref{Equ.par1}) for modelling $\DTB$. $A=1$ when $\AVXHI = 1$ and the 21-cm signal power spectrum is completely determined by density fluctuations. As reionization progresses (i.e. $\AVXHI$ deceases), $A$ first decreases and reaches its minimum value at $\AVXHI=\AVXHIMAX$. Thereafter, $A$ increases until $\AVXHI=\AVXHISTAR$ where it reaches a maximum value, which we call $A_\star$. After this, $A$ decreases and $A\rightarrow 0$ for $\AVXHI\rightarrow 0$ when reionization ends.

We model the dependence of $A$ on $\AVXHI$ in the following way. 
\begin{equation}
A=A_{\star}~\left(\frac{\AVXHI}{\AVXHISTAR}\right)^{\alpha_A}\left(\frac{1-\AVXHI}{1-\AVXHISTAR}\right)^{\beta_A} + 10^{-5\times(1-\AVXHI)}
\label{eq:A}
\end{equation}
with $\beta_A=\alpha_A ~(1-\AVXHISTAR)/\AVXHISTAR.$
In Equation (\ref{eq:A}), $\AVXHISTAR, A_\star$ and the slope $\alpha_A$ are free parameters that set the evolution of $A$ as a function of $\AVXHI$. The reionization scenarios considered in this study suggest $\AVXHISTAR\lesssim 0.5$. The first term of the right hand side in the above equation shows the $\AVXHI$ dependence of $\DTB/\Ddd$ at $k= 0.05 \hmpc$ for stages when the ionization power spectrum ($\Dxx$) dominates the 21-cm signal power spectrum in comparison with the matter density. The second part of the right hand side of this equation shows the $\AVXHI$ dependence during the initial stages of the EoR corresponding to $\AVXHI>\AVXHIMAX$ when the density power spectrum ($\Ddd$) and the anti-correlation between the density and neutral fraction ($\Dxd$) are important (see e.g., Figure \ref{image_psfid}). Note that the second term rapidly decreases as $\AVXHI$ decreases and is negligible at the stages when the peak of $A$ vs $\AVXHI$ curves occurs. Thus, we neglect the second term when we determine the values for $\beta_A$ in terms of $\alpha_A$ at the maximum. Deriving an analytical form for $\AVXHIMAX$ is not straightforward, as we estimate $\AVXHIMAX$ numerically for a set of our input parameters and thus $\AVXHIMAX$ is a derived quantity for a given model.

\begin{figure}
\begin{center}
\includegraphics[scale=0.75]{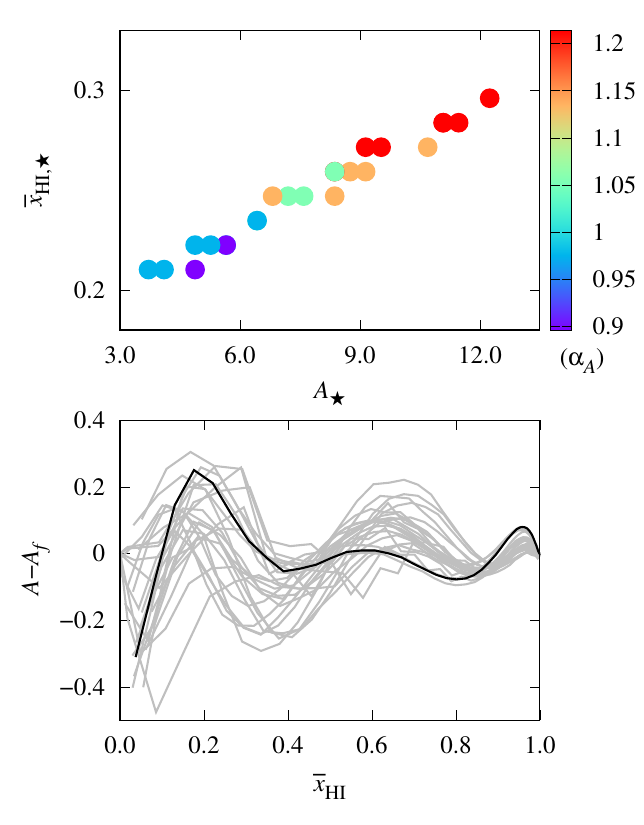}
    \caption{Results of fitting the dependence of $A$ on $\AVXHI$ using Equation (\ref{eq:A}). Top panel shows the best-fit values of the parameters $A_\star$, $\AVXHISTAR$ and $\alpha_A$ when we fit different curves of the left panel of Figure \ref{image_psfit2} using Equation (\ref{eq:A}). Here, $A$ represents the ratio $\DTB/\Ddd$ at $k\sim 0.05 \hmpc$. The bottom panel shows the difference between the input/true $A$ values and predicted/fitted $A_f$ values with the best-fit parameters as a function of average ionization fraction $\AVXHI$.}   
   \label{image_fitA}
\end{center}
\end{figure}

Next, we check the accuracy of the fitting form of $A$ as used in Equation (\ref{eq:A}). We consider the evolution of $A=\DTB/\Ddd(k=0.05 ~\hmpc)$ as a function of $\AVXHI$ from different {\sc grizzly} reionization scenarios as inputs. We vary $A_\star,\AVXHISTAR$ and $\alpha_A$ on regularly-spaced grids respectively in the ranges $[0, 20]$, $[0, 1]$ and $[0, 2]$, and determine their best-fit values. For a particular reionization scenario, we fit the evolution of $A$ values corresponding to different stages of reionization together using Equation (\ref{eq:A}). The top panel of Figure \ref{image_fitA} shows the best-fit values of the parameters $A_\star$, $\AVXHISTAR$ and $\alpha_A$. Although we observe a clear correlation between $A_\star$ and $\AVXHISTAR$, as $A$ is sensitive to small changes in these parameters, we still keep all of them as independent parameters in our final ansatz. The bottom panel of Figure \ref{image_fitA} shows the evolution of $A-A_f$ as a function of $\AVXHI$ for different {\sc grizzly} reionization scenarios. Here, $A_f$ represents the best-fit value of $A$ obtained using Equation (\ref{eq:A}).

\begin{figure}
\begin{center}
\includegraphics[scale=0.75]{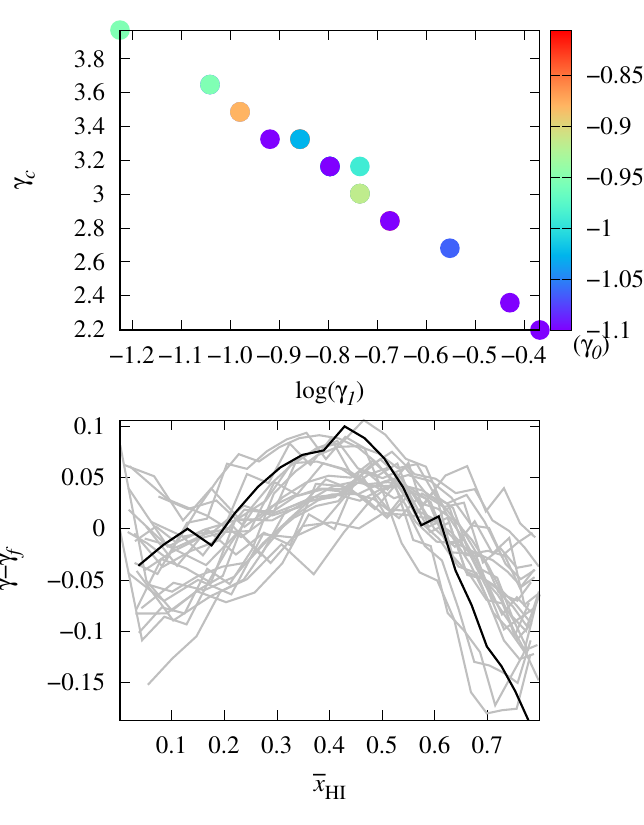}
    \caption{Results of fitting the dependence of $\gamma$ on $\AVXHI$ using Equation (\ref{Equ.gamma}). Top panel shows the best-fit values of the parameters $\log \gamma_1$, $\gamma_c$ and $\gamma_0$ when we fit different curves in the middle panel of Figure \ref{image_psfit2} using Equation (\ref{Equ.gamma}). The bottom panel shows the difference between the input/true $\gamma$ values and predicted/fitted $\gamma_f$ values with the best-fit parameters as a function of average ionization fraction $\AVXHI$. }
   \label{image_fitgamma}
\end{center}
\end{figure}

Next, we consider the $\gamma$ parameter. The middle panel of Figure \ref{image_psfit2} shows the dependencies of the best-fit values of $\gamma$ on $\AVXHI$. As expected, $\gamma\rightarrow 0$ for $\AVXHI\rightarrow 1$. For $\AVXHI > \AVXHIMAX$, $\gamma$ increases as $\AVXHI$ decreases and roughly shows a power-law dependency on $\AVXHI$. We modelled this part as $\gamma=10^{-2}~\AVXHI^{\gamma_p}$ where $\gamma_p$ is the power-law index. For $\AVXHI \lesssim \AVXHIMAX$, $\gamma$ decreases with $\AVXHI$ and 
 roughly shows an exponential drop while $\gamma$ reaches negative values for $\AVXHI\rightarrow 0$. We modelled this drop of $\gamma$ with the decreases of $\AVXHI$ as  $\gamma=\gamma_1~\exp[\gamma_c~\AVXHI] + \gamma_0$. The parameter $\gamma_c$ controls the rate of decrease of $\gamma$ for $\AVXHI \lesssim \AVXHIMAX$, while $\gamma_0+\gamma_1$ represents $\gamma$ values when $\AVXHI\rightarrow 0$. Therefore we choose the fitting form for $\gamma$ to be written as 
\begin{equation}
\begin{aligned}
\gamma & = \left(\gamma_1~\exp[\gamma_c~\AVXHI] + \gamma_0 \right) ~(1-\mathcal{H}(\AVXHIMAX))\\
& +   10^{-2}~\AVXHI^{\gamma_p} ~\mathcal{H}(\AVXHIMAX),
  \end{aligned}
\label{Equ.gamma}
\end{equation}
where, $\mathcal{H}$ is the Heaviside step function. In order to reduce the number of parameters, we use a boundary condition for $\gamma$. As the two discontinuous functional forms of $\gamma$ for $\AVXHI \le \AVXHIMAX$ and $\AVXHI > \AVXHIMAX$ have the same value for $\AVXHI=\AVXHIMAX$, we can determine $\gamma_p$ as $\gamma_p=\log{\left(\gamma_1 ~\exp[\gamma_c~\AVXHIMAX] + \gamma_0 \right)}/(0.01 ~\log{\AVXHIMAX})$.

Next, we check the accuracy of the fitting form of $\gamma$ as defined in Equation (\ref{Equ.gamma}). We consider different evolutions of $\gamma$ as a function of $\AVXHI$ from the middle panel of Figure \ref{image_psfit2}. We vary $\gamma_0, \log \gamma_1$ and $\gamma_c$ on regularly-spaced grids respectively in the ranges $[-1.5, 0]$, $[-3, 0]$ and $[0, 10]$ and determine their best-fit values. Note that the fitting considers different $\gamma$ values corresponding to different stages of a particular reionization history to obtain the best-fit values. The top panel of Figure \ref{image_fitgamma} shows the best-fit values of the parameters $\gamma_1$, $\gamma_c$ and $\gamma_0$. The bottom panel of Figure \ref{image_fitgamma} shows the deviation $\gamma-\gamma_f$ as a function of $\AVXHI$ for different {\sc grizzly} reionization scenarios. Here, $\gamma_f$ represents the $\gamma$ value for the best-fit value of the parameters following Equation (\ref{Equ.gamma}). Similarly to the fit for $A$, here we find a prominent correlation between $\gamma_c$ and $\gamma_1$. As $\gamma_c$ and $\log\gamma_1$ behave roughly linearly, we consider $\gamma_1=10^{(1.3-\gamma_c)/2.25}$ which reduces the number of parameters to $\gamma_c$ and $\gamma_0$. Here,  $\gamma_c$ accounts for the change in $k$-dependence of the bias $\DTB/\Ddd$ with $\AVXHI$ for $\AVXHI\le \AVXHIMAX$.  $\gamma_0$ accounts for the power-law dependence on $k$ feature of $\DTB/\Ddd$ in addition to small-scale feature  
$1/[1+(k/0.3)^{1.5}]$ at stages when $\AVXHI \rightarrow 0$.

\begin{table*}
\caption[]{List and description of the eight parameters used to model the redshift evolution of the EoR 21-cm brightness temperature power spectrum $\DTB$.}
\centering
\small
\tabcolsep 12pt
\renewcommand\arraystretch{1.5}
   \begin{tabular}{| c | c  |}
\hline
\hline
Parameters & Description  \\

\hline
\hline
$z_0$ & Redshift corresponding to $\AVXHI=0.5$		\\
\hline
$ \Delta z$ & Redshift range of reionization in a $tanh$ model. 	\\
\hline
$ \alpha_0$ & Asymmetry parameter around $\AVXHI=0.5$ in the redshift evolution of $\AVXHI$. 	\\
\hline
 $ A_\star$ & 	Maximum value of the bias at $k=0.05 \hmpc$.	\\
 \hline
 $\AVXHISTAR$ & Mean neutral fraction at the redshift when the bias at $k=0.05 \hmpc$ gets the maxima. 	\\
 \hline
$\alpha_A$ & Power-law index on $\AVXHI$ which accounts for the change of bias as a function of $\AVXHI$ at $k=0.05 \hmpc$.	\\
 \hline
 $\gamma_c$ &  Account for the change in scale-dependence of bias with $\AVXHI$. 		\\
 \hline
 $\gamma_0$ & 	Account for the all-scale feature 
 of bias in addition to small-scale feature  
$1/[1+(k/0.3)^{1.5}]$ at stages with $\AVXHI \rightarrow 0$. \\

\hline
\hline
\end{tabular}
\tablefoot{Here, bias represents the ratio of $\TB$ and density power spectra $\DTB/\Ddd$.}
\label{tab0}
\end{table*}

Now, we are left with an ansatz which depends on five redshift-independent parameters $A_\star,~ \AVXHISTAR,~ \alpha_A,~ \gamma_c,~{\rm and}~ \gamma_0$ to model the EoR 21-cm power spectrum as a function of $k$ modes and the reionization history, which is parametrized by the globally-averaged neutral fraction $\AVXHI(z)$. In the following section, we will introduce our modelling of reionization history.

\subsection{Modelling the reionization history}
\label{sec:xhiimodel}
The EoR 21-cm signal observations with radio interferometers are initially going to produce power spectrum $\DTB(z,k)$ at different redshifts $z$. However, our ansatz (Equations \ref{Equ.par1}, \ref{eq:A} and \ref{Equ.gamma}) predicts $\DTB$ as a function of $k$ and $\AVXHI$. Therefore, it is necessary to model the reionization history $\AVXHI$ as a function of redshift $z$.

\begin{figure}
\begin{center}
\includegraphics[scale=0.7]{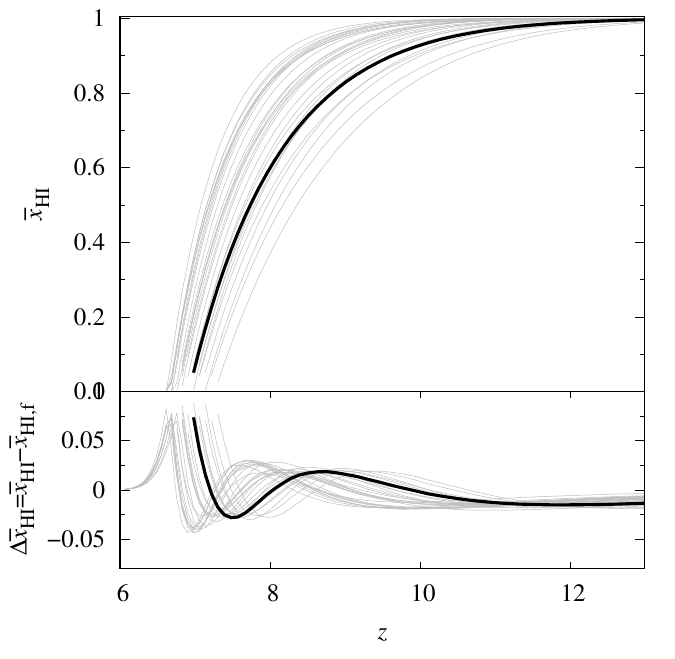}
    \caption{Modelling of reionization history $\AVXHI(z)$ using Equation (\ref{eq.xhi}). The top panel shows the evolution of the neutral fraction as a function of redshift for different reionization scenarios generated using {\sc grizzly} code. The bold black curve corresponds to our {\sc grizzly} fiducial reionization scenario. The bottom panel shows fitting error $\Delta\AVXHI=\AVXHI-\overline{x}_{\rm HI,f}$ on the redshift evolution of $\AVXHI$. We have used an asymmetric $tanh$ function (see Equation \ref{eq.xhi}) to model the redshift evolution of $\AVXHI$. }
   \label{image_xhiifit}
\end{center}
\end{figure}

The top panel of Figure \ref{image_xhiifit} shows the redshift evolution of the volume-averaged neutral fraction $\AVXHI$ for all the {\sc grizzly} reionization scenarios considered in this work. We find $\AVXHI \approx 1$ for $z\gtrsim 11$ while the bulk of reionization occurs within a narrow redshift window ($\Delta z\lesssim 4$)  at $z \lesssim 11$. Note that the reionization histories are asymmetric around $\AVXHI=0.5$. There is no well-established analytical form which accurately represents the redshift evolution of $\AVXHI$. One possible analytical form to represent the asymmetric redshift evolution of $\AVXHI$ is \citep[e.g.,][]{2017PhRvD..95b3513H}

\begin{equation}
  \begin{aligned}
    \AVXHI(z_0,\alpha_0, \Delta z, z) &= \frac{1}{2}[ 1-\tanh\{\frac{y(z_0) - y(z)}{\Delta y}\}],\\
    {\rm where} \qquad y(z) & = (1+z)^{\alpha_0}, \\
    {\rm and} \qquad \Delta y &= \alpha_0 (1+z)^{\alpha_0 -1}\times \Delta z.
\end{aligned}
\label{eq.xhi}
\end{equation}
Here $z_0$, $\Delta z$  and $\alpha_{0}$ are free parameters which govern the evolution of the mean neutral fraction $\AVXHI$ with $z_0$ representing the redshift at which $\AVXHI=0.5$, $\Delta z$ the duration  of the reionization and $\alpha_{0}$ the asymmetry of reionization history around $z_0$. Note that the parametrization of $\AVXHI$ as a function of redshift is not unique. Here, we have used an analytically simpler form for $\AVXHI(z)$. An alternative parametrization of the reionization history such as the one used in \citet{2018ApJ...858L..11T} could also produce a good fit to the reionization scenarios used in this study.

The bottom panel of Figure \ref{image_xhiifit} shows the fitting error $\Delta\AVXHI=\left(\AVXHI-\overline{x}_{\rm HI,f}\right)$ between the simulated ($\AVXHI$) and best-fit ($\overline{x}_{\rm HI,f}$) values of the reionization histories. The best-fit ionization history for a given simulated reionization scenario is obtained by varying  $z_0$, $\Delta z$  and $\alpha_{0}$ on uniformly-spaced grids and minimizing the mean square error 
\begin{equation}
    E_{\rm tot}(z_0, \Delta z, \alpha_{0}) = \frac{1}{n_z}\sum_{z<15.5} \left[\AVXHI(z_0,\alpha_0, \Delta z, z) - \AVXHI(z)\right]^2. \nonumber
\end{equation}
Here, $n_z$ is the number of redshifts considered for a reionization history, which can differ for different EoR models of {\sc grizzly}. We vary $z_0$, $\Delta z$  and $\alpha_{0}$ in the range [5, 15], [0, 3] and [0, 10] ranges, respectively. All the {\sc grizzly} reionization scenarios show a good fit with $\Delta\AVXHI \lesssim 0.03$ for the majority of the reionization redshifts ranges. At the same time, the error increases up to $\sim 0.08$ near the end of EoR when $\AVXHI\rightarrow 0$. This shows that a tanh form has some trouble capturing the fast drop  in $\AVXHI$ during the tail end of the reionization. 

In section \ref{sec:psmodelz}, we used five redshift-independent parameters to model the scale dependence of $\DTB$ as a function of $\AVXHI$. In this section, we used three $z$ independent parameters to model the redshift evolution of $\AVXHI$. Thus in the end, we are left with a set of eight redshift-independent free parameters ${\boldsymbol{\theta}} =[z_0, \Delta z, \alpha_0, A_\star, \AVXHISTAR, \alpha_A, \gamma_c, \gamma_0]$ to model both the scale dependence and redshift evolution of the EoR 21-cm power spectrum together with the reionization history. See Table \ref{tab0} for a description of the parameters.

\section{Results}
\label{sec:results}
We next apply our eight-parameter ansatz of the EoR 21-cm power spectra to various reionization scenarios obtained from different simulations. As inputs, we consider a {\sc c2ray}, a {\sc 21cmFast} and all the {\sc grizzly} simulations as mentioned in the previous section. These three 21-cm simulation frameworks are based on very different source models, but to be consistent with our ansatz for $\DTB$, we assume $\TS \gg \TCMB$ in all of them. We refer the readers to \citet{ghara15a, mellema06, 2011MNRAS.411..955M} for the details of the algorithms used in these three simulations.

\begin{figure}
\begin{center}
\includegraphics[scale=0.57]{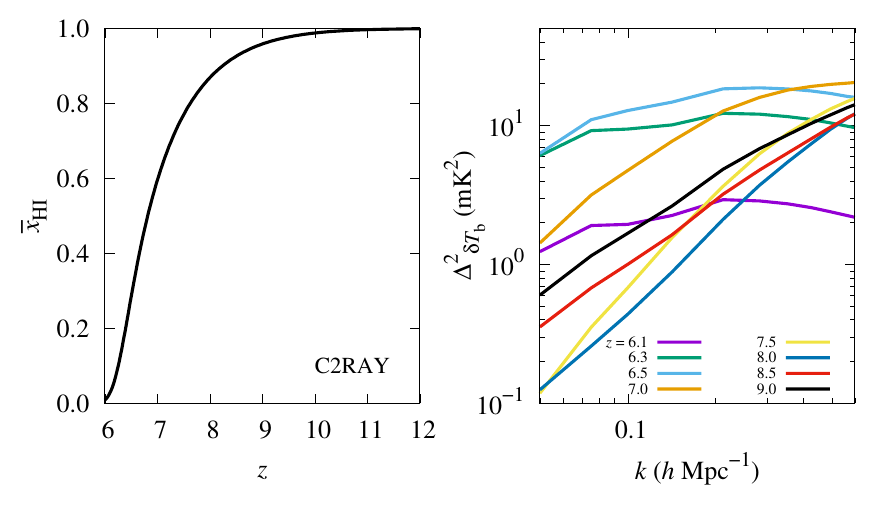}
\includegraphics[scale=0.57]{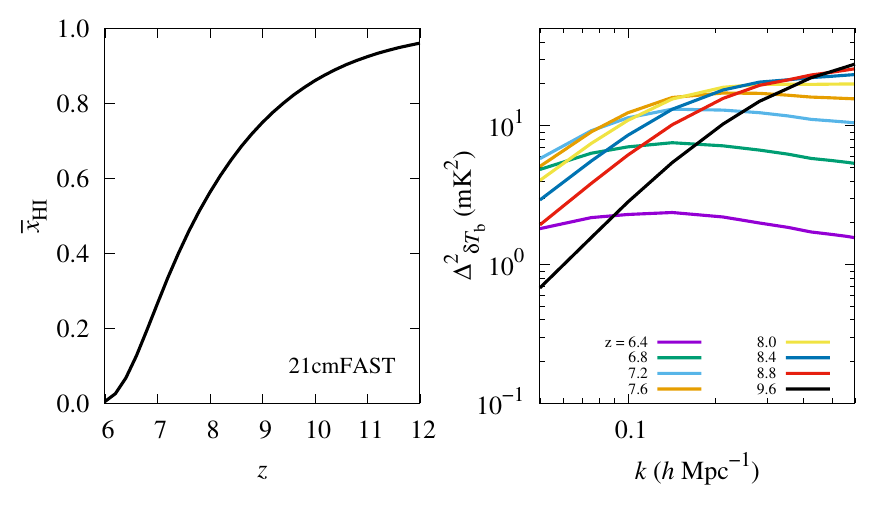}
\includegraphics[scale=0.57]{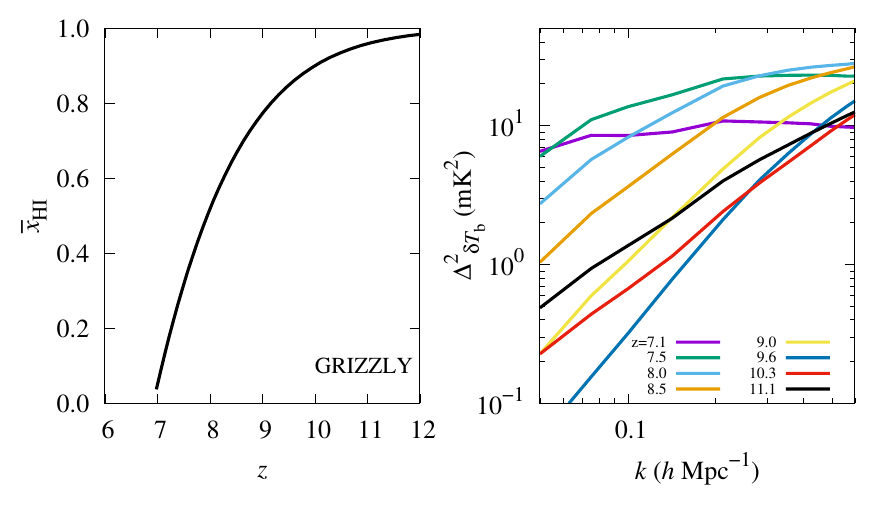}
    \caption{Three different input reionization histories and power spectra which are used to check the performance of our ansatz.  The top, middle and bottom panels correspond to {\sc c2ray}, {\sc 21cmFast} and {\sc grizzly} fiducial input reionization scenarios. The left panels show the redshift evolution of $\AVXHI$ while the right panels show the set of input power spectra used in the MCMC analysis.}
   \label{image_inputps}
\end{center}
\end{figure} 

\begin{figure*}
 \centering
\includegraphics[scale=0.45]{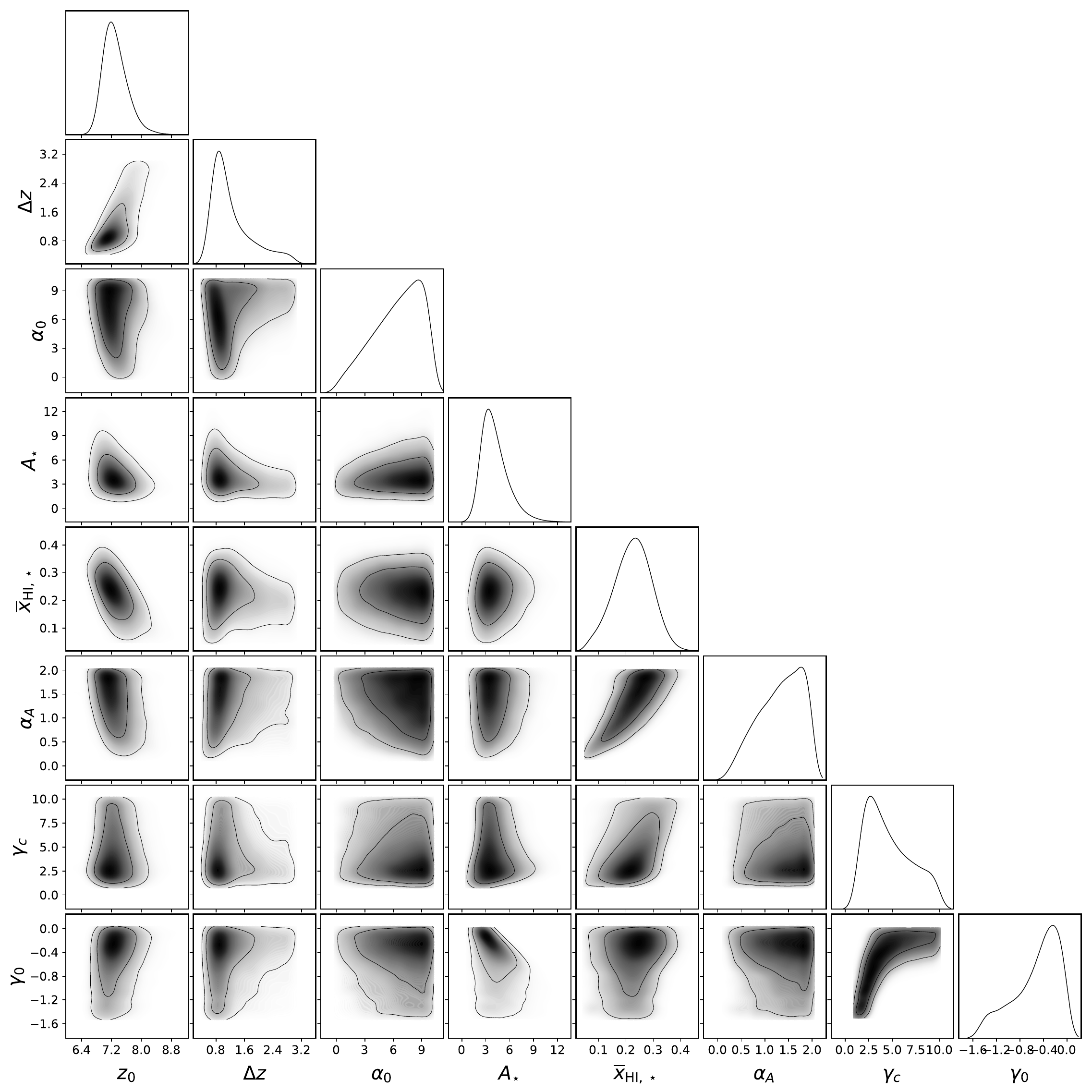}
\caption{Posterior of the ansatz parameters (see Table \ref{tab0}). This shows the constraints of the fiducial EoR scenario obtained using the MCMC analysis in terms of the EoR power spectrum model parameters. The contours in the two-dimensional contour plots represent 1$\sigma$ and 2$\sigma$ confidence levels respectively. The curves in the diagonal panels are the marginalized probability distributions of the eight parameters. The MCMC analysis is performed on inputs from {\sc c2ray}.}
   \label{image_mcmc}
\end{figure*}

\begin{figure*}
\begin{center}
\includegraphics[scale=0.47]{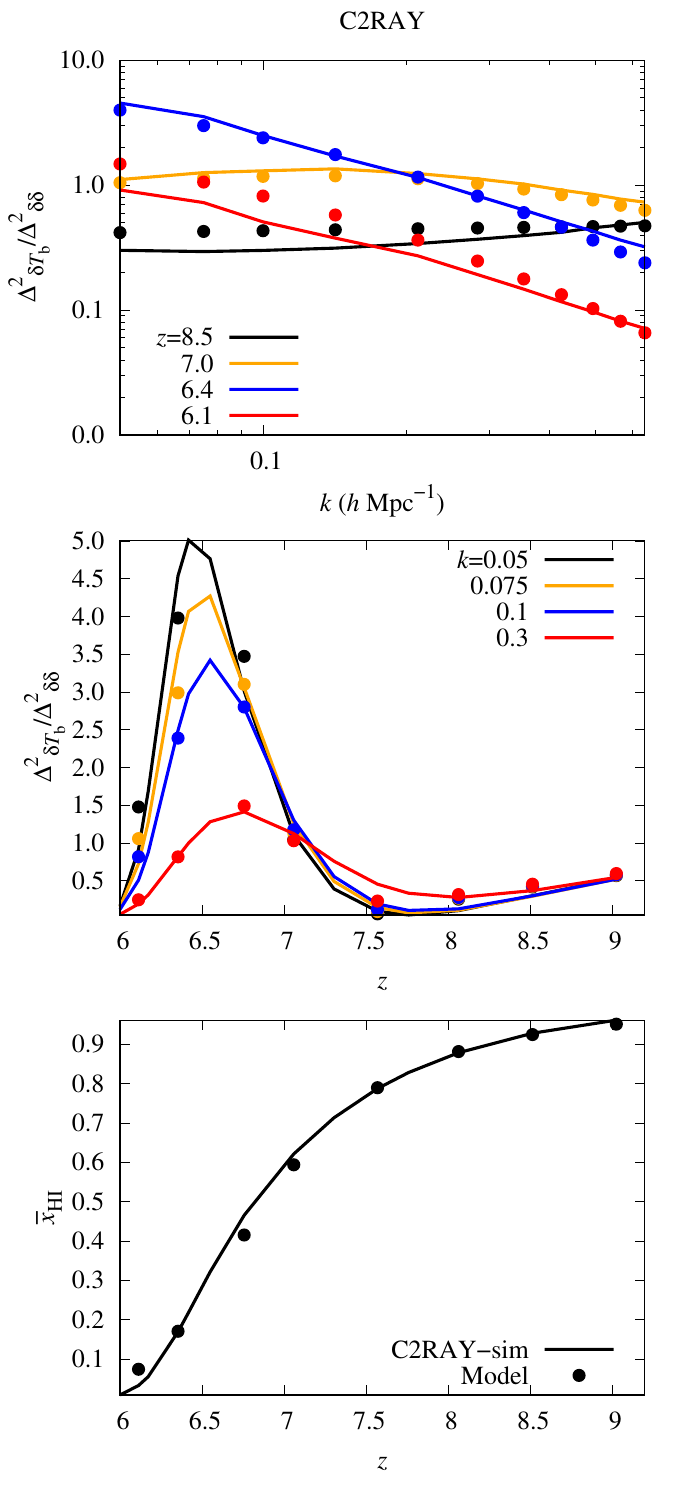}
\includegraphics[scale=0.47]{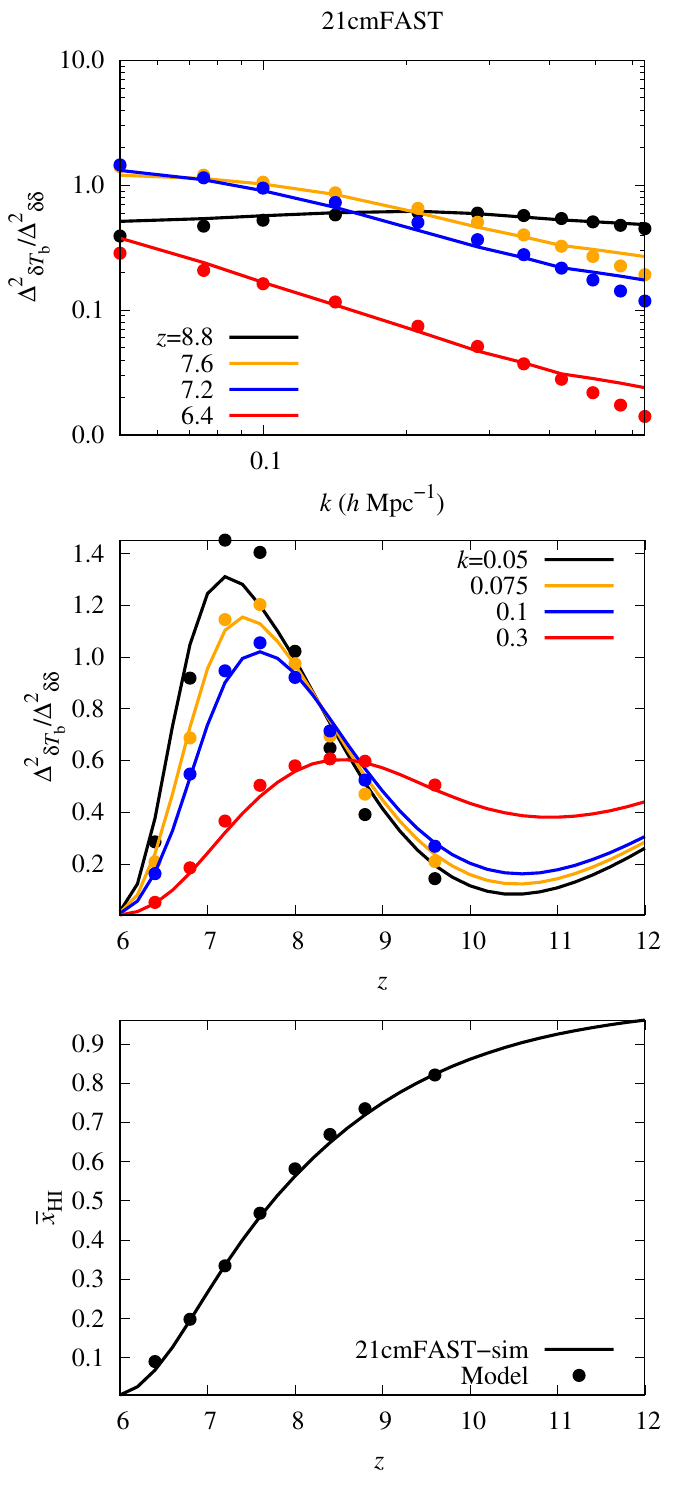}
\includegraphics[scale=0.47]{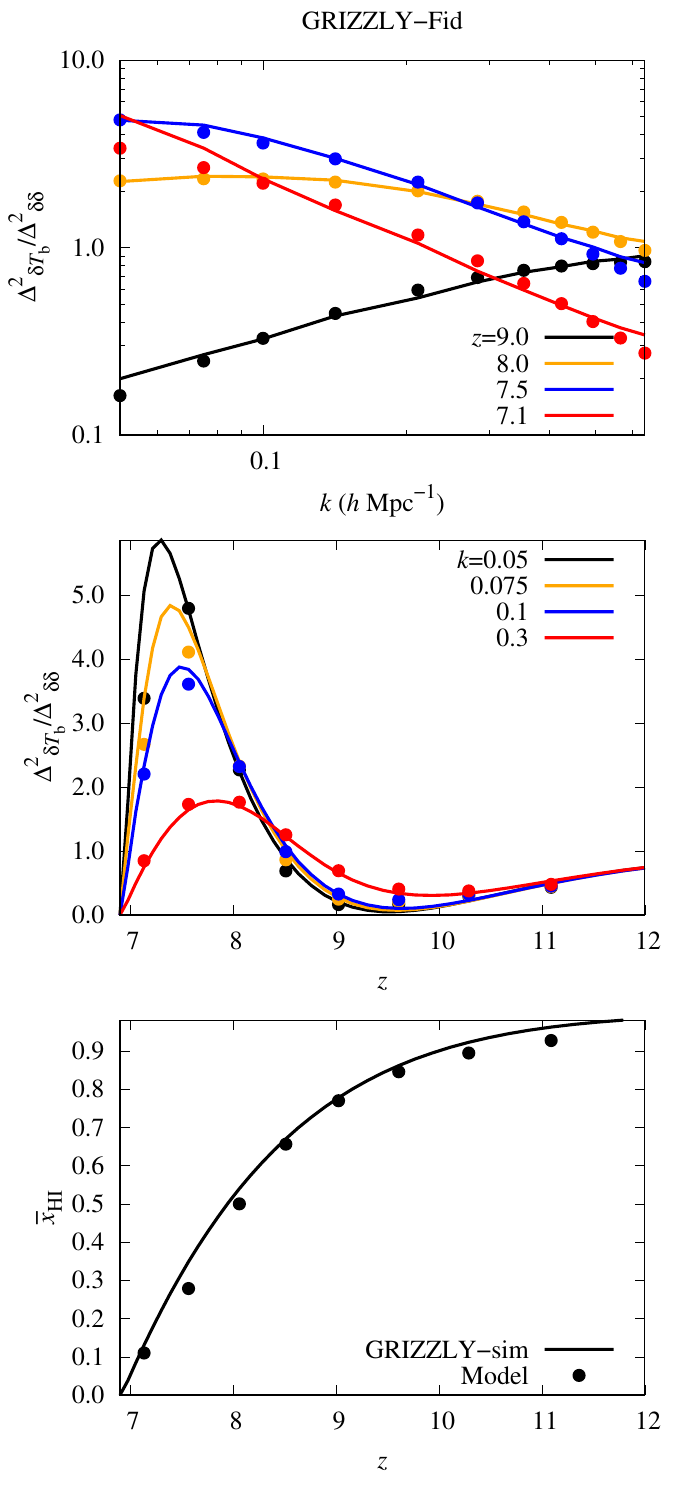}
    \caption{A comparison between the input reionization scenario and recovered model from the MCMC analysis using EoR power spectra as input to the phenomenological model considered in this work. Left to right columns represent {\sc c2ray}, {\sc 21cmfast} and {\sc grizzly} simulation respectively. From top to bottom we show, respectively, the ratio of 21-cm signal and density power spectrum ($\DTB/\Ddd$) as a function of $k-$scales at different redshifts, the redshift evolution of $\DTB/\Ddd$ for different scales and the corresponding reionization histories. The solid curves represent the inputs from simulations while the dotted curves stand for the MCMC best-fit prediction on the power spectrum model used in this work.}
   \label{image_bestfitc2ray}
\end{center}
\end{figure*}

\begin{figure}
\begin{center}
\includegraphics[scale=0.7]{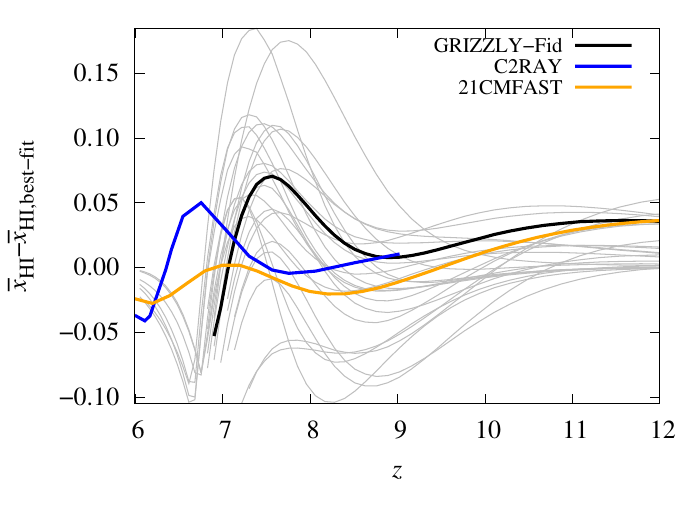}
    \caption{The redshift evolution of the difference between the input model neutral fraction and best-fit neutral fraction. The MCMC analysis here considers our eight-parameter model for the EoR power spectra. Different thin grey curves stand for different reionization scenarios from {\sc grizzly} simulation. The black curve represents our fiducial {\sc grizzly} simulation while the blue and orange curves present {\sc c2ray} and {\sc 21cmFast} input scenarios. The best-fit parameter values are obtained using input power spectra at eight redshifts for each reionization scenario. }  
   \label{image_grizzly-XHII}
\end{center}
\end{figure}

From each reionization scenario, we extract power spectra for eight different redshifts covering the majority of the EoR. The redshift ranges for {\sc c2ray}, {\sc 21cmFast} and {\sc grizzly} scenarios are $[6.1, 9]$, $[6.4, 9.6]$ and $[7.1, 11.1]$ respectively. Figure \ref{image_inputps} shows the redshift evolution of $\AVXHI$ (left column) and the corresponding simulated EoR 21-cm signal power spectra (right column) for the {\sc c2ray} (top row), {\sc 21cmFast} (middle row) and the fiducial {\sc grizzly} scenario (bottom row).  The redshifts of these power spectra span $\sim 60~{\rm MHz}$ of observational bandwidth while the frequency difference between two adjacent redshifts is $\Delta\nu\sim 7~ {\rm MHz}$. The latter is smaller than the typical bandwidth used in EoR 21-cm data analysis. For example,  \citet{2020MNRAS.493.1662M} used $12$ MHz bandwidth to estimate the upper limits on the 21-cm power spectrum at redshift $9.1$. The main motivation to use a smaller bandwidth is to resolve the fast evolving nature of the large-scale power spectrum around the peak. However, the choice of a smaller bandwidth will need more observation hours to reach the same signal-to-noise ratio.

Next we use these power spectra to constrain the eight parameters of our ansatz using Markov Chain Monte Carlo (MCMC) based parameter estimation framework. We use the publicly available code {\sc cosmomc}\footnote{\tt https://cosmologist.info/cosmomc/} \citep{2002PhRvD..66j3511L} for exploring the log-likelihood of these eight parameters $\boldsymbol{\theta}$. The log-likelihood in our MCMC algorithm is estimated as,
\begin{equation}
    \chi^2(\boldsymbol{\theta}) = - \sum_{i, j} \left(\frac{\Delta^2_{\TB,m}(z_i, k_j, \boldsymbol{\theta})-\Delta^2_{\TB,o}(z_i, k_j)}{\Delta^2_{\rm o,err}}\right)^2, \nonumber
\end{equation}
where $\Delta^2_{\TB,m}$ and $\Delta^2_{\TB,o}$ are the modelled and the simulated input power spectra, respectively. The index $i$ runs over the eight input redshifts 
while the index $j$ runs over $k-$bins with $0.05 \hmpc \leq k \leq 0.6 \hmpc$. Each MCMC analysis here is done with $8$ independent walkers (sequences of parameter values in MCMC), each of which takes $10^6$ steps.

The quantity $\Delta^2_{\rm o, \rm err}$ in the denominator is the error used in our MCMC analysis. In principle, this error should include measurement error, sample variance, and the imperfection of the ansatz. However, as the aim is to show a proof of concept of our ansatz, we consider the simple case in which the error $\Delta^2_{\rm o, \rm err}= 1 ~\rm mK^2$ is $k$ and $z$ independent. In general, $\Delta^2_{\rm o, \rm err}$ is expected to increase towards higher redshifts. Considering any observation, the scale dependence of $\Delta^2_{\rm o, \rm err}$ changes with redshift as the $uv$ coverage of an interferometric observation changes with observation frequency and also the sky noise varies with the frequency. Here, we do not consider such realistic situations and will address these issues in a follow-up work. 

\begin{table*}
\caption[]{Summary of the outputs from the MCMC analysis and constraints on the eight parameter models of the EoR power spectra.}
\centering
\small
\tabcolsep 10pt
\renewcommand\arraystretch{1.5}
   \begin{tabular}{| c | c c c c c c c |}
\hline
\hline
Scenario & Parameters & Explored  range & Mean & Standard Deviation & Best fit  & $68\%$ limits & $95\%$ limits	 \\

\hline
\hline
                & $z_0$             & [5, 15]   & 7.30   &  0.32  & 6.88    & [6.9, 7.5]	& [6.7, 7.98]		\\
                & $ \Delta z$       & [0, 3]    & 1.24   &  0.56   & 0.83  	  & [0.56, 1.40]	& [0.42, 2.50]		\\
                & $ \alpha_0$       & [0, 10]   &  6.40  &  2.49  & 6.43   	  & [5.03, 9.56]	& [1.72, 10]	\\
{\sc c2ray}     & $ A_\star$        & [0, 20]   &  4.18  &  1.61  & 5.17   	  & [2.12, 5.18]	& [1.22, 7.63]		\\
                & $\AVXHISTAR$ & [0, 1]   & 0.23   &  0.07  & 0.27  	  & [0.16, 0.30]	& [0.09, 0.35]		\\
                & $\alpha_A$        & [0.1, 2]  & 1.33   &  0.46  & 1.81     & [1.07, 1.97]	& [0.51, 2.0]		\\
                & $\gamma_c$        & [0, 10]   & 4.65   &  2.39  & 2.05	  & [1.4, 6.00]	& [1.07, 9.07]		\\
                & $\gamma_0$        & [-3, 2]   & -0.52   &  0.38  & -1.20	  & [-0.66, -0.01]	& [-1.31, 0.00]		\\
\hline
                & $z_0$             & [5, 15]   & 8.16   &  0.42  & 7.70    & [7.53, 8.70]	& [6.92, 9.46]		\\
                & $ \Delta z$       & [0, 3]    & 1.95   &  0.57   & 1.70 	  & [1.31, 2.61]	& [0.95, 2.94]		\\
                & $ \alpha_0$       & [0, 10]   &  5.91  &  2.47  & 5.50   	  & [4.01, 9.21]	& [1.56, 9.95]	\\
{\sc 21cmFast}     & $ A_\star$        & [0, 20]   &  1.39  &  0.41  & 1.60   	  & [0.86, 1.71]	& [0.59, 2.30]		\\
                & $\AVXHISTAR$ & [0, 1]   & 0.31   &  0.08  & 0.38	  & [0.23, 0.41]	& [0.13, 0.47]		\\
                & $\alpha_A$        & [0.1, 2]  & 1.27   &  0.46  & 1.90    & [0.95, 1.95]	& [0.43, 2.0]		\\
                & $\gamma_c$        & [0, 10]   & 3.78   &  1.80  & 3.50	  & [1.55, 4.62]	& [0.94, 7.51]		\\
                & $\gamma_0$        & [-3, 2]   & -0.68   &  0.43  & -0.80  & [-0.96, 0.05]	& [-1.40, 0.0]		\\
\hline
                & $z_0$             & [5, 15]   & 8.03   &  0.21  & 8.10    & [7.8, 8.2]	& [7.60, 8.44]		\\
                & $ \Delta z$       & [0, 3]    & 1.25   &  0.20   & 1.20  	  & [1.11, 1.40]	& [0.96, 1.56]		\\
                & $ \alpha_0$       & [0, 10]   &  5.61  &  2.30  & 6.60   	  & [5.1, 9.8]	& [2.0, 10]	\\
{\sc grizzly}     & $ A_\star$        & [0, 20]   &  5.21  &  0.71  & 5.20   	  & [4.41, 5.80]	& [3.83, 6.61]		\\
                & $\AVXHISTAR$ & [0, 1]   & 0.26   &  0.06  & 0.24  	  & [0.22, 0.32]	& [0.15, 0.40]		\\
                & $\alpha_A$        & [0.1, 2]  & 1.46   &  0.38  & 1.30     & [1.20, 1.94]	& [0.80, 2.0]		\\
                & $\gamma_c$        & [0, 10]   & 2.41   &  0.90  & 2.40	  & [1.60, 3.20]	& [0.76, 3.79]		\\
                & $\gamma_0$        & [-3, 2]   & -1.05   &  0.51  & -0.88	  & [-1.20, -0.50]	& [-2.50, -0.25]		\\
\hline
\hline
\end{tabular}
\tablefoot{Top-to-bottom panels stand for the {\sc c2ray}, {\sc 21cmFast} and {\sc grizzly} fiducial input scenarios. We use eight input power spectra for each of the reionization scenarios. The left to right columns represent the scenarios considered in this work, set of parameters, their ranges used in the MCMC analysis, mean parameter value, standard deviation, best-fit value, $68\%$ and $95\%$ credible limits obtained from the MCMC chains. The MCMC analysis is done using $8$ chains each with $10^6$ steps.}
\label{tab1}
\end{table*}

We run the MCMC analysis on the {\sc c2ray, 21cmfast} and {\sc grizzly} scenarios. The input power spectra for each scenario are presented in Figure \ref{image_inputps}. The outcomes of the MCMC analysis are presented in the following subsections.

\subsection{Scenario I: {\sc c2ray}}
\label{sec:c2ray}
First, we consider a reionization scenario which is generated using the EoR 21-cm signal modelling code {\sc c2ray} \citep{mellema06}. This code uses the gridded density field and halo lists from $N$-body simulations and applies `Conservative Causal Ray-tracing method' based 3D radiative transfer to produce ionization fraction fields at different stages of reionization. The set of $\TB$ power spectra and the redshift evolution of $\AVXHI$ of the input reionization history as obtained from a {\sc c2ray} simulation are shown in the top row of Figure \ref{image_inputps}. This scenario uses the same dark-matter fields and halo list as used in the {\sc grizzly} simulations. {\sc c2ray} also considers contributions from all dark-matter halos with their masses larger than $10^9 ~\MSUN$ and assumes the rate of production of the ionizing photons to be $1.3\times 10^{42} \times M_{\rm halo}/\MSUN ~\rm s^{-1}$. In this simulation, the mean-free-path length of the ionizing photons is chosen as $70~{\rm Mpc}$.  In this {\sc c2ray} reionization scenario, $\AVXHI$ decreases from $0.9$ to $0$ as reionization progressed between $z\approx 8$ and $6$. We find that the evolution of the ratio $\DTB/\Ddd$ at $k=0.05 \hmpc$ reaches a maximum value of $\approx 5$ at $z\approx 6.4$ when $\AVXHI \approx 0.3$.

Figure \ref{image_mcmc} shows the posteriors of the eight parameters $\boldsymbol{\theta}$ of our power spectrum ansatz when we use the set of {\sc c2ray} power spectra as inputs to our MCMC framework. The off-diagonal panels show the joint probability distribution for a pair of parameters where 2D contours represent $1\sigma$ and $2\sigma$ confidence levels respectively. The curves in the diagonal panels represent the marginalized probability distributions of the individual ansatz parameters. The plot shows that while most of the parameters are well-constrained, some of them are not. One reason behind this might be the degeneracy of these parameters with the other parameters. The best-fit parameter values obtained from this analysis are $z_0=6.9,~ \Delta z=0.83,~  \alpha_0=6.4,~ A_\star=5.2,~ \AVXHISTAR=0.27,~ \alpha_A=1.8,~ \gamma_c=2,~{\rm and}~ \gamma_0=-1.2$ (see also Table \ref{tab1}). The best-fit values of $A_\star$ and $\AVXHISTAR$ agree well with the input reionization scenario which corresponds to $A_\star=5.01$ and $\AVXHISTAR=0.22$. The comparison between input and best-predicted models for this reionization scenario is shown in the left column of Figure  \ref{image_bestfitc2ray}. Here The top-left panel shows the comparison as a function of $k$ at different redshifts while the middle-left panel shows the redshift evolution of the predicted and input ratio $\DTB/\Ddd$ for different $k-$bins. The curves indicate that our ansatz performs well at different stages of the reionization and for the $k$ range we considered here. A comparison between the input reionization history and the ansatz predictions (bottom-left panel of Figure \ref{image_bestfitc2ray}) suggests an excellent recovery of the redshift evolution of $\AVXHI$ using our model. Figure \ref{image_grizzly-XHII} shows the corresponding deviation  $\AVXHI -\overline{x}_{\rm HI, best-fit}$ where $\overline{x}_{\rm HI, best-fit}$ represents the prediction using the MCMC best-fit parameters. Here, we find that the deviation remains between $\pm 0.05$ as indicated by the blue thick curve.

\subsection{Scenario II: {\sc 21cmFast}}
Our second input reionization scenario is generated using the publicly available semi-numerical 21-cm code {\sc 21cmFast} \citep{2011MNRAS.411..955M, Park2019InferringSignal}. In this semi-numerical approach, the density fields are generated following the first-order perturbation theory \citep{1970A&A.....5...84Z} while the ionization fields are produced using the excursion-set approach \citep{2004ApJ...613....1F}. We assume the following parameters: fraction of galactic gas in stars for $10^{10} ~\MSUN$ halo $f_{\star, 10}=0.05$, the power-law index for star formation and halo mass relation $\alpha_\star = 0.5$, the UV ionizing escape fraction for $10^{10} ~\MSUN$ halo $f_{\rm esc, 10}=0.1$, the power-law index for UV escape fraction  and halo mass relation $\alpha_{\rm esc}=-0.5$, the characteristic mass scale for star formation suppression $M_{\rm turn}=5\times 10^8 ~\MSUN$, star-formation timescale in units of the Hubble time $t_\star=0.5$, number of ionizing photons per stellar baryon $N_\gamma=5000$, and mean-free path of ionizing photons $R_{\rm mfp}= 50$ Mpc \citep[for the details of these parameters, see][]{Park2019InferringSignal}. In this case, reionization ends at $z \sim 6$ while the majority of the ionization happens below $z\lesssim 10$. The corresponding input reionization history and the input power spectra are shown in the middle row of Figure \ref{image_inputps}. The input set of the 21-cm power spectra of the reionization scenario shows features similar to those of the fiducial {\sc grizzly} and {\sc c2ray} scenarios. We find that the ratio $\DTB/\Ddd$ at $k=0.05 \hmpc$ reaches a maximum value of $\approx 1.3$ at $z\approx 7.2$ corresponding to $\AVXHI=0.35$.

We repeat the MCMC analysis (as done for {\sc c2ray}) to obtain the posterior of our ansatz parameters. The middle row of Table \ref{tab1} shows the posterior constraints on the eight parameters of our ansatz. The best-fit parameter values are $z_0=7.7,~ \Delta z=1.7,~ \alpha_0=5.5,~ A_\star=1.6,~ \AVXHISTAR=0.38,~ \alpha_A=1.9,~ \gamma_c=3.5,~{\rm and}~\gamma_0=-0.8$. The corresponding power spectra predictions at all the redshifts agree well with the input simulated power spectra (see the top and middle panels of the central column of Figure \ref{image_bestfitc2ray}). Similar to the previous case, the best-fit values of $A_\star$ and $\AVXHISTAR$ agree well with the input {\sc 21cmfast} reionization scenario which has $A_\star=1.3$ and $\AVXHISTAR=0.34$. We compare the simulated reionization history $\AVXHI(z)$ with that predicted from our MCMC analysis in the bottom-central panel of Figure \ref{image_bestfitc2ray}. Similar to the {\sc c2ray} scenario, our framework works efficiently in this case as well and recovers reionization history $\AVXHI$ within a maximum deviation between $\pm 0.03$ as shown by the orange line in Figure \ref{image_grizzly-XHII}.

\subsection{Scenario III: {\sc grizzly}}
The simulated input power spectra and the reionization history for the fiducial {\sc grizzly} reionization scenario are shown in the bottom panels of Figure \ref{image_inputps}. As described in section \ref{sec:psmodelXHI}, this corresponds to {\sc grizzly} parameters $\zeta=1$, $M_{\rm min}=10^9 ~\MSUN$ and $\alpha_S=1$ \citep[for details, see][]{2020MNRAS.493.4728G}. This results in a reionization history where the IGM gets $10\%$ ionized at $z\approx 11$ while the reionization ends around $z\approx 7$ (see top right panel of Figure \ref{image_bestfitc2ray}). Here, the ratio $\DTB/\Ddd$ at $k=0.05 \hmpc$ reaches a maximum value of $\approx 6$ at $z \approx 7.3$ which corresponds to $\AVXHI=0.25$.

The posterior constraints on the ansatz parameters, obtained using MCMC analysis, are summarized in the bottom row of Table \ref{tab1}. The best-fit parameter values are $z_0=8.1,~ \Delta z=1.2,~ \alpha_0=6.6,~ A_\star=5.2,~ \AVXHISTAR=0.24,~ \alpha_A=1.3,~ \gamma_c=2.4,~{\rm and}~ \gamma_0=-0.88$. Similar to the previous cases, the predicted power spectra and the best-fit values of $A_\star$ and $\AVXHISTAR$ are in good agreement with the simulated input values. The $A_\star$ and $\AVXHISTAR$ values from the input reionization scenario are $5.8$ and $0.22$, respectively. The predicted evolution of $\AVXHI$ as shown in the top right panel of Figure \ref{image_bestfitc2ray} shows good agreement within the absolute deviations of $0.05$.

We repeat our MCMC analysis for the other $23$ {\sc grizzly} models as represented by the shaded grey lines in Figures \ref{image_xhiifit} and \ref{image_psevo} considering the input power spectra at the same redshifts used for the fiducial {\sc grizzly} model. We plot the difference between the input and predicted neutral fraction for all these models in Figure \ref{image_grizzly-XHII}. These suggest that our ansatz represents the EoR 21-cm power spectrum close enough and can recover the reionization history $\AVXHI(z)$ within an absolute error of $\sim 0.1$ (see the grey lines in Figure \ref{image_grizzly-XHII}). Note that the evolution of $\AVXHI$ as a function of redshift is very steep around $z_0$ and thus even a small error in the estimate of $z_0$ will result in a large difference between the input(true) and predicted $\AVXHI$. This is evident from the plots for various {\sc grizzly} models in Figure \ref{image_grizzly-XHII} between redshift $7$ and $8$. The deviations could become larger for the models having steeper slopes around $z_0$.

\section{Summary \& Conclusions}
\label{sec:con}

The redshifted 21-cm signal from the IGM during the EoR encodes unique information about that period. The 21-cm observations indirectly probe the properties of the ionizing and heating sources and directly probes the ionization and thermal states of the IGM during the first billion years of our Universe. In this study, we focus on inferring the properties of the EoR IGM, rather than those of astrophysical sources, through 21-cm signal observations. The large-scale amplitude and scale-dependent features of the EoR 21-cm brightness temperature power spectrum depend on the ionization fraction of hydrogen and the morphology and distribution of the ionized/neutral regions. Our main aim is to develop a source parameter-free phenomenological model that constrains the properties of the EoR IGM using multi-redshift 21-cm power spectrum measurements. The framework constrains the redshift evolution of the average neutral fraction and a set of quantities related to the morphology and distribution of the ionized regions.

Using different {\sc grizzly} simulations, we study the scale-dependent features of the 21-cm power spectra at different stages of EoR. The quantity we aim to model is the ratio of the 21-cm brightness temperature ($\TB$) and the density power spectrum also known as the 21-cm bias in the literature.  We modelled this ratio as $\DTB/\Ddd = A \frac{\left(\frac{k}{0.05}\right)^\gamma}{1+\left(\frac{k}{0.3}\right)^{1.5}}$. 
Here, $A$ represents bias at $k=0.05 \hmpc$.  We tested the goodness of fit of our ansatz at various stages of reionization using $24$ different {\sc grizzly} scenarios. These tests suggest that the aforementioned functional form of the ratio of $\TB$ and density power spectra efficiently reproduce the EoR 21-cm power spectra for different reionization histories, accurately within $\lesssim10\%$ error (see Figure \ref{image_perrorsfit}).

As the $A$ and $\gamma$ parameters in the above-mentioned ansatz evolve during reionization, we additionally model how these parameters evolve as a function of $\AVXHI$. The model for $A$ (Equation \ref{eq:A}) uses three parameters, the maximum value of the ratio ($ A_\star$), the corresponding neutral fraction $\AVXHISTAR$ and a power-law index $\alpha_A$. The evolution of $\gamma$ can be described with two parameters (Equation \ref{Equ.gamma}) :  $\gamma_c$ which accounts for the change in scale-dependence, and 
 $\gamma_0$ which accounts for the deviation of the scale dependence of $\DTB/\Ddd$ from 
$1/[1+(k/0.3)^{1.5}]$ at small-scales (see section \ref{sec:psmodelz} for details).

Using the {\sc grizzly} simulations, we fit the evolution of $\AVXHI$ as a function of redshift using three parameters (see Equation \ref{eq.xhi}). These are: redshift $z_0$ which corresponds to $\AVXHI=0.5$, redshift range of reionization $\Delta z$ in a $\tanh$ reionization model and asymmetry parameter $\alpha_0$ to invoke asymmetry in history around $\AVXHI=0.5$. We tested the goodness of this form of $\AVXHI(z)$ using $24$ different {\sc grizzly} reionization models and found them to be consistent within $\Delta \AVXHI=\pm 0.1$ (see Figure \ref{image_xhiifit}).

In the end, we are left with a set of eight redshift and scale-independent parameters  $\boldsymbol{\theta} =[z_0, \Delta z, \alpha_0, A_\star, \AVXHISTAR, \alpha_A, \gamma_c, \gamma_0]$ to jointly model the redshift evolution and scale-dependence of the ratio $\DTB/\Ddd$. We demonstrate the performance of this ansatz with $24$ {\sc grizzly} models, one reionization scenario from {\sc c2ray} and one from {\sc 21cmFast}. We use as an input the power spectra simulated at eight redshifts within the interval $z=[7.1-11.1]$ (corresponds to $\sim 60~ {\rm MHz}$ bandwidth) and perform a Bayesian MCMC analysis to constrain the ansatz parameters for each of the three reionization scenarios. 
All these tests collectively indicate that our ansatz reproduces the scale-dependence and the redshift evolution of the ratio $\DTB/\Ddd$ reasonably well for a variety of reionization models considered here. The predicted redshift evolution of $\AVXHI$, using the best-fit MCMC parameter set $\boldsymbol{\theta}$, matches nicely with the input reionization history within $\Delta \AVXHI=\pm 0.1$ (see Figure \ref{image_grizzly-XHII}). At the same time the constrained values for $A_\star, \AVXHISTAR$ match closely with the reionization scenarios (see Figure \ref{image_bestfitc2ray} and Table \ref{tab1}).

Our aforementioned approach is similar in spirit to a few previous studies. For example, \citet{battaglia13} simulate a reionization redshift field for a given density field (filtered at a particular scale). However, their approach fundamentally assumes a strong correlation between the density and reionization redshift fields. On the other hand, \citet{2018JCAP...10..016M} take a perturbative approach to construct an EoR 21-cm signal field using the underlying density field. Their formalism assumes some source field and patchiness-dependent bias factors as well as a characteristic size of ionized regions to connect the 21-cm signal field with the density field. Unlike them, our approach is conceptually simpler and directly connects the EoR 21-cm signal power spectrum with the matter power spectrum. This phenomenological model directly exploits the features of multi-redshift EoR 21-cm signal power spectra for predicting the quantities related to the EoR IGM states. Thus, it makes our ansatz more flexible, computationally inexpensive and faster while exploring IGM parameters in the context of current and future observations. Additionally, our phenomenological ansatz is agnostic to the various methods of EoR simulations (3D \& 1D radiative transfer and excursion-set based), whereas the analysis in \citet{2022MNRAS.514.2010M} is restricted to the excursion-set based simulations only.

The analysis presented in this paper is only based on the `inside-out' reionization scenario where the highly dense regions around the radiating sources become ionized first. We speculate that the same is also applicable for an `outside-in' reionization model as the scale-dependence of the EoR power spectra as well as the redshift evolution of the large-scale power spectrum are qualitatively similar to the `inside-out' case for the wavenumber range of our interest \citep[see e.g., Figure 2 and 3 of][]{2020MNRAS.498..373P}. However, in the `outside-in' case, we do not expect a trough of the large-scale power spectra at $\AVXHIMAX$ and thus we expect an even simpler form of the ansatz for the EoR power spectra. We leave a detailed investigation for the `outside-in' reionization scenario for a future study.

The accuracy of the ansatz predictions of the EoR 21-cm power spectra suggests that it will be useful to constrain reionization scenarios using existing and upcoming measurements from LOFAR, MWA, HERA, and SKA. Here, we have used a simple-minded constant error on the input power spectra. It will be interesting to see the performance of this model when realistic errors are taken into account. We plan to address this in our future work.

Our model is based on $\TB$ power spectra with high spin temperatures. This model also works if the gas temperature of the neutral IGM remains uniform. However, things get complicated when the presence of spin temperature fluctuations modifies the power spectra significantly. One expects a more complex evolution of the power spectrum in that case. Modelling such behaviours is out of the scope of this paper and will be addressed in a follow-up work.

Here we consider only the EoR 21-cm power spectra measurements to constrain the reionization history. In addition, information from several other probes such as the Thomson scattering optical depth measurement from CMB observation \citep[e.g.][]{2020A&A...641A...6P} and the Gunn-Peterson trough in high-$z$ quasar spectra \citep[e.g.][]{Fan06b, Becker_2015},  observations of high-$z$ Ly-$\alpha$ emitters \citep[e.g.][]{Hu10, Morales_2021}, and the Ly-$\alpha$ damping wings in high-$z$ quasar spectra \citep[e.g.][]{2018Natur.553..473B} can also be combined with the \HI measurements from the EoR to tighten the constrain on the reionization history. We plan to address this in a future study using {\sc polar} \citep{2023MNRAS.522.3284M} algorithm which is based on {\sc grizzly} and the semi-analytical galaxy formation code L-Galaxies 2020 and self-consistently model the evolution of galaxy properties during the EoR.

Although our framework is efficient in recovering the redshift evolution of the average ionization fraction using the measurements of the EoR 21-cm signal power spectrum, there are several aspects that need improvements. While most of the parameters of this framework are easy to interpret, understanding the physical meaning of a few parameters such as $\gamma_c$ and $\gamma_0$ is non-trivial. These parameters are connected to the morphology and distribution of the ionized regions. It is important to understand how these quantities are linked to the distribution of morphological quantities such as volume, surface and mean curvature \citep[see e.g.,][]{2018MNRAS.479.5596G, 2021MNRAS.505.1863G, 2024MNRAS.530..191G}. This study is also beyond the scope of this work. We plan to address it in a future study.

\begin{acknowledgements}
RG, AKS and SZ acknowledge support from the Israel Science Foundation (grant no. 255/18). Furthermore, RG acknowledges support from the Kaufman Foundation (Gift no. GF01364). SZ also acknowledges Alexander von Humboldt Foundation for the Humboldt Research award. AKS also acknowledges support from National Science Foundation (grant no. 2206602). GM acknowledges support by the Swedish Research Council grant 2020-04691.  LVEK acknowledges the financial support from the European Research Council (ERC) under the European Union's Horizon 2020 research and innovation programme (Grant agreement No. 884760, ``CoDEX'').
\end{acknowledgements}

\bibliographystyle{aa}
\bibliography{mybib}

\label{lastpage}
\end{document}